\documentclass[12pt,preprint]{aastex}
\usepackage{epsfig}
\usepackage{natbib}
\usepackage{graphicx}
\usepackage{slashbox}
\usepackage{multirow}
\usepackage{lscape}
\usepackage{mathrsfs,amssymb}
\usepackage{amsmath}
\usepackage{subfigure}
\usepackage{amssymb}
\usepackage{multirow}
\usepackage{tabularx}
\usepackage{rotating}
\usepackage{amsmath}
\usepackage{ulem}

\newcommand       \cm           {\,{\rm cm}}

\newcommand       \K            {\,{\rm K}}

\newcommand       \yr       {\,{\rm yr}}

\newcommand       \NH           {N_{\rm H}}
\newcommand       \simlt        {\lesssim}
\newcommand       \simgt        {\gtrsim}

\newcommand       \mum          {\,{\rm \mu m}}
\newcommand       \ppm          {\,{\rm ppm}}

\newcommand       \simali       {\sim\,}

\newcommand \dprim {\left[\rm D/H\right]_{\rm prim}}
\newcommand \Dism {\left[\rm D/H\right]_{\rm ISM}}
\newcommand \dism {\left[\rm D/H\right]_{\rm ISM}}

\newcommand \dgas {\left[\rm D/H\right]_{\rm gas}}

\newcommand \Dpah {\left[\rm D/H\right]_{\scriptscriptstyle\rm PAH}}

\newcommand       \Iratio         {I_{4.4}/I_{3.3}}
\newcommand       \Aratio        {A_{4.4}/A_{3.3}}
\newcommand       \Acd           {A_{4.4}}
\newcommand       \ACD           {A_{4.4}}
\newcommand       \Aaro          {A_{3.3}}
\newcommand       \ACH          {A_{3.3}}
\newcommand       \NC         {N_{\rm C}}
\newcommand       \ND         {{N_{\rm D}}}
\newcommand       \Iratioobs     {\left(I_{4.4}/I_{3.3}\right)_{\rm obs}}

\newcommand       \NCH         {N_{\rm C-H}}
\newcommand       \NCD         {{N_{\rm C-D}}}
\newcommand       \km        {\,{\rm km}}
\newcommand       \Liter      {\,{\rm L}}
\newcommand       \mol       {\,{\rm mol}}
\newcommand       \mole       {\,{\rm molecule}}

\newcommand       \varepsilonWV {\varepsilon_{\tilde{\nu}}}
\newcommand       \varepsilonmax {\varepsilon_{\tilde{\nu},{\rm max}}}

\countdef\decade=200
\advance\decade by \year
\countdef\hours=201
\advance\hours by \time
\divide\hours by 60
\countdef\mins=202
\advance\mins by \hours
\multiply\mins by 60
\multiply\hours by 100
\countdef\miltime=203
\advance\miltime by \hours
\advance\miltime by \time
\advance\miltime by -\mins
\def\today{\number\decade.\number\month.\number\day.\number\miltime}

\shorttitle{IR spectra of Deuterated PAHs}
\title{
Deuterated Polycyclic Aromatic Hydrocarbons
in the Interstellar Medium:
The C--D Band Strengths of Mono-Deuterated Species
\\{\small DRAFT: \today ~~}
}
\author{X.J.~Yang\altaffilmark{1,2},
            Aigen Li\altaffilmark{2},
            and R.~Glaser\altaffilmark{3}}
\altaffiltext{1}{Department of Physics,
                      Xiangtan University,
                      411105 Xiangtan, Hunan Province, China;
                      {\sf xjyang@xtu.edu.cn}}
\altaffiltext{2} {Department of Physics and Astronomy,
                  University of Missouri,
                  Columbia, MO 65211, USA;
                  {\sf lia@missouri.edu}}
\altaffiltext{3} {Department of Chemistry,
                  Missouri University of Science and Technology,
                  Rolla, MO 65409, USA;
                  {\sf glaserr@mst.edu}}

\begin{document}

\begin{abstract}
Deuterium (D) is one of the light elements
created in the big bang.
As the Galaxy evolves, the D/H abundance
in the interstellar medium (ISM)
decreases from its primordial value due to ``astration''.
However, the observed gas-phase D/H abundances
of some sightlines in the local Galactic ISM
are substantially lower than the expected reduction
by astration. The missing D could have been depleted
onto polycyclic aromatic hydrocarbon (PAH) molecules
which are ubiquitous and abundant in interstellar regions.
%
To quantitatively explore the hypothesis of PAHs
as a possible reservoir of interstellar D,
we compute quantum-chemically the infrared vibrational
spectra of mono-deuterated PAHs and their cations.
We find that, as expected,
when H in PAHs is replaced by D,
the C--H stretching and bending modes
at 3.3, 8.6 and 11.3$\mum$ shift to longer wavelengths
at $\simali$4.4, 11.4 and 15.4$\mum$, respectively,
by a factor of $\simali$$\sqrt{13/7}$, the difference
in reduced mass between the C--H and C--D oscillators.
We derive from the computed spectra
the mean intrinsic band strengths of
the 3.3$\mum$ C--H stretch
and 4.4$\mum$ C--D stretch
to be $\langle\Aaro\rangle\approx13.2\km\mol^{-1}$
and $\langle\ACD\rangle\approx7.3\km\mol^{-1}$
for neutral deuterated PAHs
which would dominate the interstellar
emission at 3.3 and 4.4$\mum$.
%
By comparing the computationally-derived
mean band-strength ratio
of $\langle\Acd/\Aaro\rangle\approx 0.56$ for
neutral PAHs with the mean ratio of
the observed intensities of
$\langle\Iratio\rangle\approx 0.019$,
we find that the degree of deuteration
(i.e., the fraction of peripheral atoms
attached to C atoms in the form of D)
is $\simali$2.4\%,
corresponding to a D-enrichment
of a factor of $\simali$1200
with respect to the interstellar D/H abundance.
%
%
\end{abstract}


\section{Introduction\label{sec:intro}}
Deuterium (D) is a primordial element that was
only made during the first minutes after the
Big Bang (Epstein 1976).
The primordial D/H abundance ($\dprim$)
established by the Big Bang Nucleosynthesis (BBN)
depends sensitively on the cosmological parameters
(e.g., the baryon closure parameter,
$\Omega_b$, and the Hubble constant, $h$;
see Boesgaard \& Steigman 1983).
D could be easily destroyed by nuclear fushion
in stellar interiors,
a process known as ``astration''.
Through astration, D is converted to $^3$He, $^4$He,
and heavier elements and as a result,
the cosmic D/H abundance is expected
to decrease monotonically along with
the chemical evolution of the Galaxy
(Mazzitelli \& Moertti 1980).
Therefore, the interstellar D/H abundance
in the present epoch is directly related to
the primordial nucleosynthesis
and the subsequent Galactic chemical evolution.

To reliably measure $\dprim$, one often relies on metal-poor
quasar absorption line systems or damped Ly$\alpha$
absorption (DLA) systems which are compositionally
as ``pristine'' as possible and thus still retain
a primordial composition of D.
To date, the primordial D/H abundance has been reliably
determined to be $\dprim$\,$\approx$\,25--28$\ppm$,
based on high precision measurements
of the column densities of D\,I and H\,I
of seven such systems
(e.g., see Cooke et al.\ 2018, Zavarygin et al.\ 2018).
Alternatively, one can also infer $\dprim$ from
$\eta$, the ratio of the baryons to photons,
which is simply related to $\Omega_b h^2$
(Burles et al.\ 2001).
In the standard BBN model,
$\eta$ is the single parameter that predicts
the primordial light element abundances
including $\dprim$
(e.g., see Boesgaard \& Steigman 1983).
Using the $\Omega_b$ and $h$ parameters
derived from observations of
the cosmic microwave background,
Spergel et al.\ (2003), Coc et al.\ (2004), and
Sanchez et al.\ (2006) found $\dprim$\,$\approx$\,26$\ppm$,
which is in close agreement with that measured from
the pristine quasar absorption systems.
%
As the Galaxy evolves, D is ``astrated''
and the D/H abundance
in the interstellar medium (ISM)
decreases from its primordial value.
This is because D is destroyed in the Galaxy
as gas is cycled through stars.
On the other hand, infall to the disk of the Galaxy
of less processed or astrated gas,
which is D rich and metal poor,
from the intergalactic medium (IGM)
and small galaxies captured by the Milky Way Galaxy,
would raise the interstellar D/H abundance
closer to the primordial value.
Careful studies of the competition
between stellar destruction and infall derived
the present day interstellar D/H abundance
to be $\dism$\,$\simgt$\,$20\pm1\ppm$
(e.g., see Prodanovic et al.\ 2010).

Accurate measurements of the gas-phase
D/H abundance have been made
for more than 40 interstellar lines of sight,
based on the high resolution ultraviolet (UV)
absorption spectra  of H\,I and D\,I,
obtained with  the {\it Copernicus} satellite,
the {\it International Ultraviolet Explorer}
(IUE) satellite, and
the {\it Far Ultraviolet Spectroscopic Explorer}
(FUSE) satellite,
as well as
the {\it Goddard High Resolution Spectrometer} (GHRS)
and {\it Space Telescope Imaging Spectrograph} (STIS)
instruments aboard
the {\it Hubble Space Telescope} (HST), and
the {\it Interstellar Medium Absorption Profile Spectrograph }
(IMAPS) on the ORFEUS-SPAS II mission
(see Wood et al.\ 2004, Draine 2006, Linsky et al.\ 2006
and references therein).
These measurements revealed an unexplained mystery:
the gas-phase D/H abundance varies significantly by
a factor of $\simali$4 from one sightline to another
within a few hundred parsecs of the Sun
(see, e.g., Moos et al.\ 2002; Steigman 2003;
Wood et al.\ 2004; H\'ebrard 2005).

It had been suggested that the observed
regional variations in $\dgas$ could be
due to the regional variations in astration.
The gas returned to the ISM from stars is
expected to be nearly devoid of D,
as the D is converted to $^3$He
during the pre-main-sequence evolution
of stars with a mass of $M\simlt5M_\odot$
(Mazzitelli \& Moretti 1980).
As a result, the D/H abundance in the ISM
is an indicator of the fraction of
the baryons now in the ISM that have
passed through a star.
If different sightlines have different
star formation histories so that the interstellar
matter has been cycled through stars
to different degrees, then the D astration
will be different for different sightlines:
for sightlines with a stronger star formation activity,
more D will be destroyed and therefore one would
expect a smaller $\dgas$.
However, it is difficult to imagine how regions
separated by only a few hundred pc can have had
extremely different star formation histories.
It is also difficult to imagine why turbulent diffusion
has not homogenized the ISM in regions
situated just a few hundred pc apart.
Turbulent mixing is expected to be effective
at mixing gas over length scales of hundreds
of pc on $10^9\yr$ time scales
and one would expect that the elemental abundances
in the gas have been homogenized (see Draine 2006).
%

As early as 1982, it had been suggested
that D might be depleted from the gas phase
and sequestered in dust grains (Jura 1982).
Allamandola et al.\ (1987) analyzed the carbonaceous
matter in interplanetary dust particles (IDPs) and meteorites
and found that they are deuterium enriched
(i.e., the D/H ratios are considerably greater
than the canonical interstellar D/H ratio).
Sandford et al.\ (2001) have revealed
several processes that could lead to
deuterium enrichment in carbonaceous material
(also see Wiersma et al.\ 2020).
%
Draine (2004) argued that carbonaceous
interstellar grains could incorporate
enough D atoms to substantially reduce
the D abundance in the gas phase
and the observed variations in the gas-phase
$\dgas$ abundance could be attributed to
variations from one sightline to another
in the fraction of the D sequestered
in dust grains.

As the smallest carbonaceous grains,
polycyclic aromatic hydrocarbon (PAH)
molecules could be a major reservoir
of interstellar D (Draine 2004).
PAHs are abundant and widespread
throughout the Universe, as revealed by their
distinctive set of infrared (IR) emission bands
at 3.3, 6.2, 7.7, 8.6, 11.3 and 12.7$\mum$,
which are characteristic of their vibrational
modes (see Li 2020).
PAHs of intermediate size are expected
to become deuterium enriched in space
through the selective loss of hydrogen
during photodissociation events.
Since the aromatic C--D bond zero-point energy
is about 30\% lower than that of the C--H bond,
hydrogen loss is favoured over deuterium loss.
For PAHs in typical interstellar environments,
the C--D bond rupture rate is $\simali$3 times
lower than that of C--H bond,
the PAH D/H ratio might therefore be substantially
larger than the interstellar abundance of
$\Dism\sim2\times10^{-5}$
(Allamandola et al.\ 1989).
Draine (2006) showed that collisions of
D with PAH cations, and collisions of
D$^+$ with PAHs, are expected to result
in incorporation of D into PAHs,
with the rate of such collisions
sufficiently rapid to deplete D
from the gas on time scales of $\simali$2\,Myr
in cool diffuse clouds.

PAHs show three emission features associated with
vibrations of the C--H bond:
the C--H stretch at 3.3$\mum$,
the in-plane C-H bending at 8.6$\mum$, and
the out-of-plane (oop) C--H bending at 11.3$\mum$.
The stretching and bending modes can be approximated
as harmonic oscillators. If a peripheral H atom is
replaced by D, the reduced mass of the C--H oscillator
will increase by a factor 13/7, and the wavelength ($\lambda$)
will be increased by a factor $\simali$$\sqrt{13/7}$.
Therefore, deuterated PAHs will emit at
$\simali$4.4$\mum$ (C--D stretching),
$\simali$11.7$\mum$ (C--D in-plane bending),
and $\simali$15.4$\mum$
(C--D oop bending).
While the 11.7$\mum$ C--D feature
would be confused with
the C--H oop bending modes
at $\simali$11.3$\mum$
and the 15.4$\mum$ C--D feature
falls in a region where other PAH features
(i.e., the C--C--C skeleton modes) are present,
the 4.4$\mum$ C--D feature appears
to be the best probe of the deuteration
of interstellar PAHs
(see Hudgins et al.\ 1994, 2004,
Bauschlicher et al.\ 1997, Draine 2006).

Observationally, the C--D emission feature
at $\simali$4.4$\mum$ of interstellar
deuterated PAHs is often much weaker than
the C--H features at 3.3$\mum$
(e.g., see Peeters et al.\ 2004,
Onaka et al.\ 2014, Doney et al.\ 2016).
The detection and identification of
the feature as due to C--D are hampered
by its small intensity which would put it
at the limit of modern observational techniques,
including the {\it Short Wavelength Spectrometer} (SWS)
on board the {\it Infrared Space Observatory} (ISO)
and the {\it Infrared Camera} (IRC)
on board {\it AKARI}.\footnote{%
   The {\it Infrared Spectrograph} (IRS)
   on board {\it Spitzer} only operates
   longward of $\simali$5.2$\mum$.
   }
This will change with the launch of
{\it James Webb Space Telescope} (JWST).
The {\it Near InfraRed Spectrograph} (NIRSpec)
instrument on {\it JWST} will cover
the wavelength range of the C--H and C--D stretches
of deuterated PAHs with high spectral resolution
and unprecedented sensitivity.
{\it JWST}'s unique high sensitivity and
high resolution near-IR capabilities will
open up an IR window unexplored by
{\it Spitzer} and unmatched by {\it ISO} observations
and thus will place the detection of deuterated PAHs
on firm ground and enable far more detailed band
analysis than previously possible,
and therefore will allow us to quantitatively
test the deuterium depletion hypothesis
as a viable explanation of the observed
regional variations of the gas-phase
D/H abundance by accurately determining
the deuteration degree of interstellar PAHs.

To this end, one requires a prerequisite knowledge
of the intrinsic strengths of the C--D bands of
deuterated PAHs. This motivates us to compute
the IR spectra of astrophysically relevant
deuterated neutral and ionized PAHs.
Our goal is to computationally determine
the intrinsic strengths of the C--D bands.
To our knowledge, laboratory IR spectra have
only been obtained from a number of matrix-isolated
perdeuterated PAHs (i.e., PAHs in which all of
the peripheral H atoms have been replaced by D atoms;
Hudgins et al.\ 1994, Bauschlicher et al.\ 1997).
However, as mentioned earlier, while the interstellar PAH
population is expected to be enriched in D
over the canonical interstellar D/H ratio,
it is highly unlikely that they would be fully deuterated
(see Hudgins et al.\ 2004).
Moreover, the absolute C--D band intensities have not been
measured for those matrix-isolated perdeuterated PAHs
(Hudgins et al.\ 1994, Bauschlicher et al.\ 1997).
On the other hand, quantum chemical computations
have so far been limited to perdeuterated PAHs
(Bauschlicher et al.\ 1997),
D-enriched superhydrogenated PAHs
(i.e., PAHs with extra H and D atoms
attached to some of their peripheral C atoms;
Hudgins et al.\ 2004),
deuteronated PAHs
(i.e., PAHs to which a deuteron is added;
Buragohain et al.\ 2015),
deuterated ovalene (Buragohain et al.\ 2016),
and PAHs with D-substitued alphatic sidegroups
(Buragohain et al.\ 2020).
%
%
%
%
In this work, we will systematically calculate
the IR spectra of all the isomers of
a large range of singly deuterated PAH species
and thoroughly explore the effects of deuteration
on their IR spectra. Multi-deuterated PAHs will be
the focus of  a separate, companion paper.
We will obtain the C--D band intensities of
deuterated PAHs and apply them to astronomical data
to derive the deuteration degrees---the fractions
of peripheral (H and D) atoms attached to C atoms
in the form of D---of interstellar PAHs.

This paper is organized as follows.
In \S\ref{sec:Method} we briefly describe
the computational methods and
the selected target molecules.
The computed IR vibrational spectra
and the derived intrinsic C--D band strengths
are reported in \S\ref{sec:Results}.
In \S\ref{sec:discussion} we apply
the derived band strengths to
ISO/SWS and AKARI/IRC data
to determine the deuteration degrees
of interstellar PAHs.
Finally, we summarize our major results
in \S\ref{sec:summary}.

\section{Computational Methods
         and Target Molecules
         \label{sec:Method}
         }
\begin{figure*}
\centering{
\includegraphics[scale=0.65,clip]{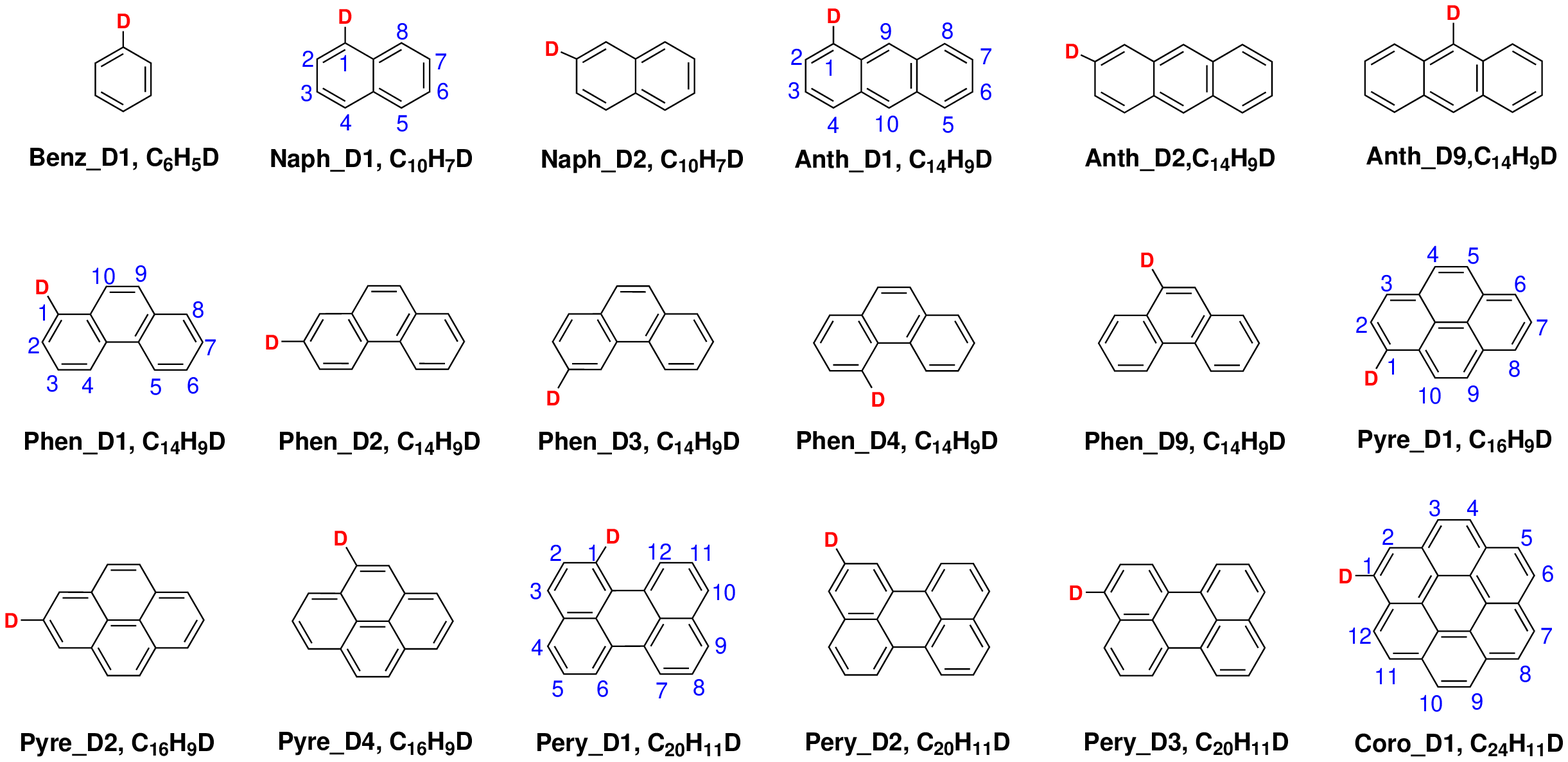}
}
\caption{\footnotesize
         \label{fig:1D_PAH_structures}
         Structures of mono-deuterated PAHs.
         We refer to a mono-deuterated species
         by the abbreviation of the first four letters of
         the name of its parent molecule
         followed by the position where the D atom
         is attached (e.g., Anth$\_$D2 refers to anthracene
         with the D atom attached at position 2).
         }
\end{figure*}

\begin{figure*}
\centering{
\includegraphics[scale=0.75,clip]{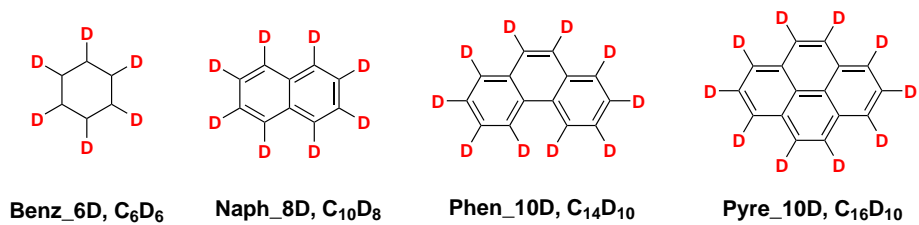}
}
\caption{\footnotesize
         \label{fig:perD_Pyrene_structures}
         Structures of fully-deuterated (i.e., perdeuterated) PAHs.
         }
\end{figure*}

We use the Gaussian09 software (Frisch et al.\ 2009)
to calculate the IR vibrational spectra for a series of
deuterated PAH molecules and cations of various
numbers of carbon atoms ($\NC$).
Our target molecules include all the possible
singly deuterated isomers of an ensemble of
seven PAH species in both neutral and cationic forms
(see Figure~\ref{fig:1D_PAH_structures}).\footnote{%
  Here we focus on small PAHs of $\NC\simlt24$
  since deuterium enrichment in PAHs of $\NC\simgt40$
  is not expected: these larger PAHs
  have a sufficiently large number of internal degrees
  of freedom to accommodate the maximum energy
  of typical UV photons without subsequent photolytic
  bond cleavage occurring (Hudgins et al.\ 2004).
  %
  }
 %
%
To facilitate comparison with experimental results,
we will also calculate four perdeuterated species
(see Figure~\ref{fig:perD_Pyrene_structures}):
perdeuterated benzene (C$_6$D$_6$),
naphthalene (C$_{10}$D$_8$),
phenanthrene (C$_{14}$D$_{10}$)
and pyrene (C$_{16}$D$_{10}$).
We note here that in our target molecules,
all the D atoms are attached to the aromatic C
atoms, therefore our calculations will mainly explore
the vibrational properties of aromatic C--D bonds.

We will refer mono-deuterated species
by the abbreviation of the first four letters of
the names of their parental PAH molecules
followed by the position where the D atom
is attached (e.g., Naph$\_$D2 refers to
naphthalene with the D atom attached at position 2).
%

We employ the hybrid density functional theoretical
method (B3LYP) at the {\rm 6-311+G$^{\ast\ast}$} level,
which gives sufficient calculational accuracies
with operable computer time
(see Yang et al.\ 2017 and reference therein).
The standard scaling is applied to
the frequencies by employing
a scale factor of $\simali$0.9688
(Borowski 2012).
%

\section{Results\label{sec:Results}}
\subsection{Perdeuterated PAHs
                    \label{subsec:PerdeuteratedPAH_Results}}
We first show our calculational results
for perdeuterated PAHs, because these are
the ones that experimental data are available
for comparison. Our target molecules are
perdeuterated benzene (Benz$\_$6D),
naphthalene (Naph$\_$8D), phenanthrene (Phen$\_$10D)
and pyrene~(Pyre$\_$10D). The experimental spectrum
of Benz$\_$6D is taken from
the {\it National Institute of Standards and Technology}
(NIST) webbook, while the others are taken
from Bauschlicher et al.\ (1997).
%
Since the experimental data do not have information
for the band intensities, we will only compare
the relative intensities with our calculations.

\begin{figure*}
\centering{
\includegraphics[scale=0.5,clip]{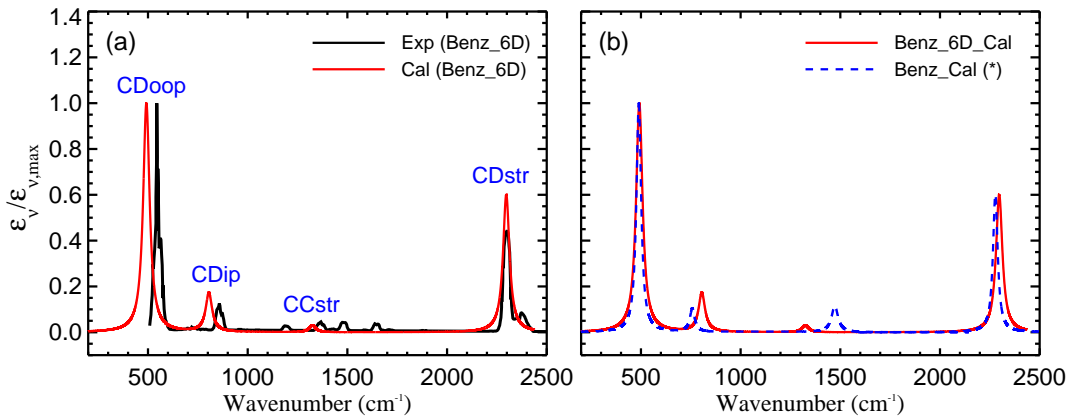}
}
\caption{\footnotesize
         \label{fig:PerDeu_Benzene}
         Left panel (a):
         Comparison of the experimental spectrum
         of fully-deuterated (i.e., perdeuterated)
         benzene taken from
         the NIST webbook (black solid line)
         with that computed from B3LYP/6-311+G$^{\ast\ast}$
         (red solid line).
         The C--D bands are labeled as
         CDstr (for the C--D stretching bands),
         CDip (for the C--D in-plane bending bands),
         and CDoop (for the C--D oop bending bands).
         Also labeled (as CCstr) is the C--C stretching bands.
         To better match the experimental spectrum,
         the calculated frequencies are scaled with
         a factor of 0.9748 (instead of the canonical
         scale factor of 0.9688) and a line width
         of 20$\cm^{-1}$ is assigned.
          To facilitate comparison, we normalize
          the experimental and computational spectra
          (in terms of molar extinction coefficient $\varepsilonWV$)
          to their maxima ($\varepsilonmax$)
          since the experimental spectrum
          does not have information for the (absolute) intensities.
          Right panel (b):
          Comparison of the computational spectrum
          of fully-deuterated benzene (red solid line)
          with that of pure benzene (blue dashed line).
          For pure benzene, the C--H bands originally
          at $\simali$663$\cm^{-1}$ (oop bending),
          1026$\cm^{-1}$ (in-plane bending)
          and 3085$\cm^{-1}$ (stretching)
          are scaled with a factor of $\sqrt{7/13}$
          while the features related to the benzene ring
          at $\simali$1463$\cm^{-1}$ stay intact.
          }
\end{figure*}

\begin{figure*}
\centering{
\includegraphics[scale=0.5,clip]{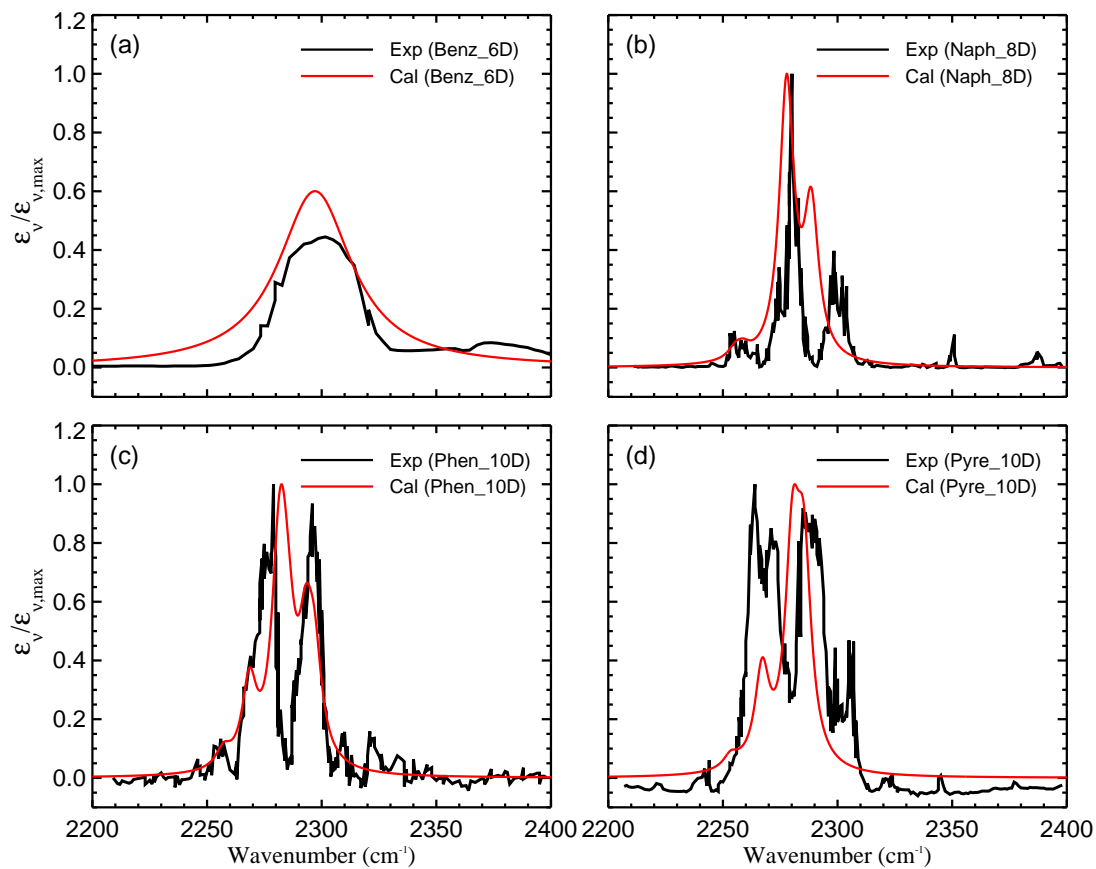}
}
\caption{\footnotesize
         \label{fig:PerDeu_spec_com}
         Comparison of the experimental spectra
         in the C--D stretching region
         (black solid lines) of perdeuterated benzene (a),
         naphthalene (b), phenanthrene (c) and
         pyrene (d) with those computed from
         B3LYP/6-311+G$^{\ast\ast}$ (red solid lines).
         Panel (a) is a zoomed-in view of
         the C--D stretch of  perdeuterated benzene
         shown in Figure~\ref{fig:PerDeu_Benzene}.
         While the experimental spectrum of perdeuterated
         benzene is taken from the NIST webbook,
         the experimental spectra of the other three
         perdeuterated species are taken from
         Bauschlicher et al.\ (1997).
         The calculated frequencies are scaled with
         a factor of 0.9688 and a line width
         of 4$\cm^{-1}$ is assigned for perdeuterated
         naphthalene, phenanthrene and
         pyrene, while a scale factor of 0.9748
         and a line width of 20$\cm^{-1}$
         are applied to benzene.
          To facilitate comparison, we normalize
          the experimental and computational spectra
          to their maxima since the experimental spectra
          do not have information for the (absolute) intensities.
          }
\end{figure*}

Figure~\ref{fig:PerDeu_Benzene}a shows
the computed IR spectrum of Benz$\_$6D (red line),
expressed as the molar extinction coefficient
$\varepsilonWV$ normalized to its peak value
$\varepsilonmax$,
where $\tilde{\nu}\equiv\lambda^{-1}$
is the wavenumber.\footnote{%
   The molar extinction coefficient $\varepsilonWV$
   measures how strongly a chemical species
   or substance absorbs light at a particular wavelength.
   It is an intrinsic property of chemical species
   that is dependent upon their chemical composition
   and structure.
   The absorption cross section per molecule
   $C_{\rm abs}(\lambda)$ relates to $\varepsilonWV$
   through
   \begin{equation}
    \label{eq:epsilon2Cabs}
    \frac{C_{\rm abs}(\lambda)}
            {\cm^2\mole^{-1}}
    = \frac{1}{6.02\times10^{16}}\times
        \frac{\varepsilonWV}
                {\Liter\mol^{-1}\cm^{-1}} ~~.
   \end{equation}
   The band strength $A_\lambda$ relates to
    $\varepsilonWV$ and $C_{\rm abs}(\lambda)$
    through
   \begin{equation}
    \label{eq:epsilon2A}
   \frac{A_\lambda}{\km\mol^{-1}}
   = 100\times\int  \frac{\varepsilonWV}
              {\Liter\mol^{-1}\cm^{-1}}\,d\tilde{\nu}
   = 6.02\times10^{18}\times
       \int \frac{C_{\rm abs}(\lambda)}
            {\cm^2\mole^{-1}}\,d\tilde{\nu} ~~,
   \end{equation}
   where the integrations are performed
   over the band.
   }
The experimental spectrum of Benz$\_$6D
taken from the NIST webbook is also shown
in Figure~\ref{fig:PerDeu_spec_com}a (black line).
By applying a single scaling factor of 0.9748
(instead of the canonical scaling factor of 0.9688)
for the calculated frequencies,
the computational C--D stretching feature
closely matches that of the experimental spectrum,
while the other features at smaller wavenumbers
seems to require a somewhat larger scaling factor.

Benz$\_$6D is highly symmetric, and its vibrational
modes are highly degenerate just as benzene.
Its IR spectrum shows only four peaks,
which is quite similar to benzene.
The only difference is that the C--D bond has
a larger by a factor of 13/7 reduced mass,
therefore the vibrational modes related to
the C--D bond will appear at a smaller by a
factor of $\simali$$\sqrt{7/13}$ wavenumber.
To illustrate, we show in Figure~\ref{fig:PerDeu_Benzene}b
the IR spectrum of benzene calculated
with the same method (blue dashed line).
We scale the three features related to
the C--H vibrations at wavenumbers
of $\simali$663$\cm^{-1}$ (C--H oop bending),
1026$\cm^{-1}$ (C--H in-plane bending)
and 3085$\cm^{-1}$ (C--H stretching)
with a factor of $\sqrt{7/13}$,
while keeping the benzene-ring-related
features at $\simali$1463$\cm^{-1}$ intact.
Figure~\ref{fig:PerDeu_Benzene}b clearly demonstrates
that after scaling, the frequencies of the C--D and C--H
stretching and oop bending features match quite
well with each other, while the C--D in-plane bending
features of perdeuterated PAHs are blueshifted
and relatively weaker.
Meanwhile, when the H atoms are replaced with D,
the C--C stretching features at $\simali$1460$\cm^{-1}$
are redshifted and seriously depressed.

Figure~\ref{fig:PerDeu_spec_com} shows
the calculated and experimental spectra of
Benz$\_$6D (a), Naph$\_$8D (b), Phen$\_$10D (c)
and Pyre$\_$10D (d) in the frequency range of 
2200--2400$\cm^{-1}$.
Here we focus on the C--D stretching bands
of Naph$\_$8D, Phen$\_$10D and Pyre$\_$10D
(Figure~\ref{fig:PerDeu_spec_com}a
is a zoomed-in view of
Figure~\ref{fig:PerDeu_Benzene}a).
The experimental spectra
basically show two major peaks
and several satellite features.
The calculated frequencies match the experimental
ones quite well after scaling with the factor of 0.9688
(0.9748 for Benz$\_$6D)
and applying a line width of 4$\cm^{-1}$
(20$\cm^{-1}$ for Benz$\_$6D).
Again, since only the relative intensities were measured
for the C--D bands from the experimental spectra
(Bauschlicher et al.\ 1997), we compare the relative
band intensities of the experimental spectra
with those of our calculations.
As tabulated in Table~\ref{tab:Freq_Int_Comp_CalExp},
the relative intensities are quite consistent with each other.
In view of the close match between our calculations and the
experimental data, we conclude that our calculations
provide reliable frequencies and relative intensities
for the C--D vibrational modes.

\subsection{Mono-Deuterated PAHs
                    \label{subsec:MonoD_Results}}
\begin{figure*}
\centering{
\includegraphics[scale=0.5,clip]{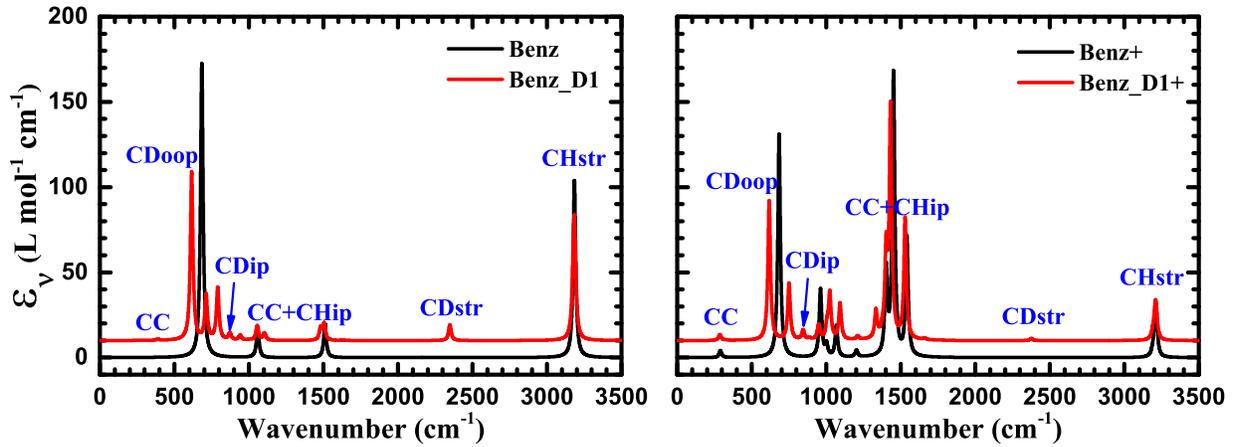}
}
\caption{\footnotesize
         \label{fig:Benz_IR_spec_all}
         Comparison of the computational spectra (red lines)
         of mono-deuterated benzene (left panel)
         and its cation (right panel)
         with that of pure benzene and its cation (black lines).
         The frequencies are scaled with
         a factor of 0.9688, and a line width
         of 10$\cm^{-1}$ is assigned.
         The computational spectra are expressed
         in terms of molar extinction coefficient $\varepsilonWV$
         which relates to the absorption cross sections $C_{\rm abs}(\lambda)$
         through eq.\,\ref{eq:epsilon2Cabs}.
         The major vibrational bands are labeled as
         CDstr, CDip, CDoop, CHstr, and CHip
         (C--H in-plane bending bands).
         }
\end{figure*}

\begin{figure*}
\centering{
\includegraphics[scale=0.5,clip]{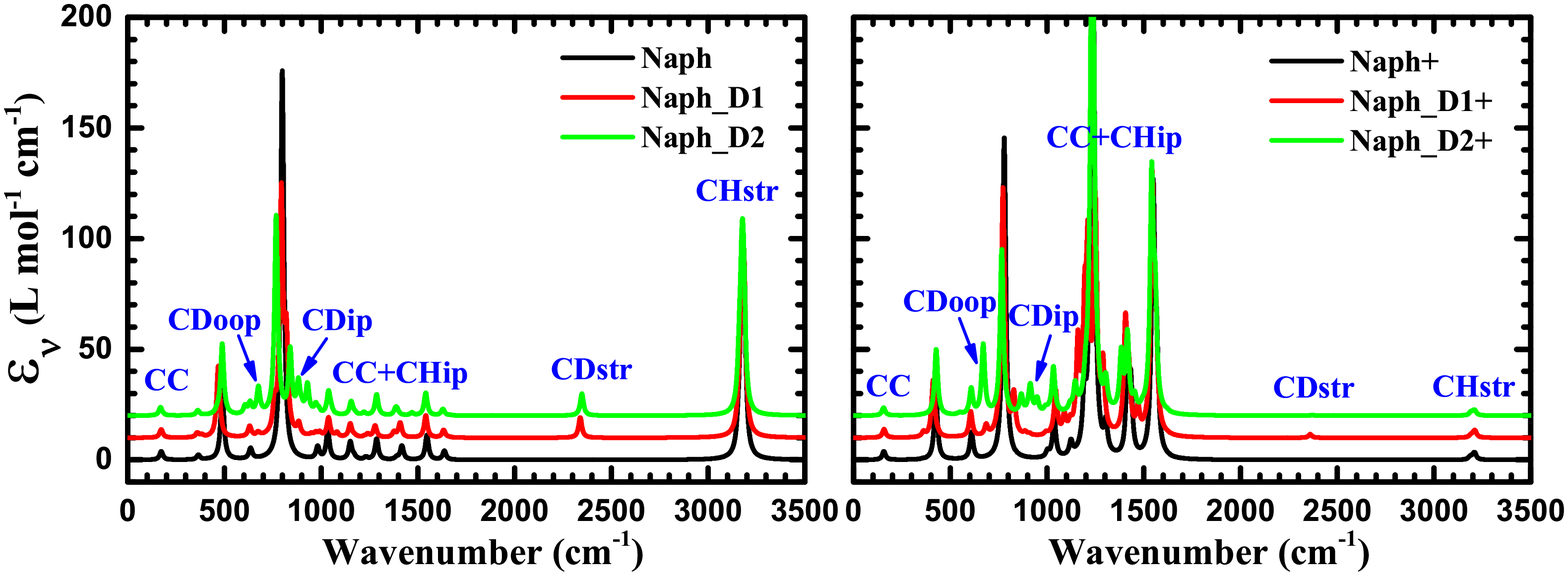}
}
\caption{\footnotesize
         \label{fig:Naph_IR_spec_all}
         Comparison of the computational spectra
         (red and green lines)
         of mono-deuterated naphthalene (left panel)
         and its cation (right panel)
         with that of pure naphthalene and its cation (black lines).
         The frequencies are scaled with
         a factor of 0.9688, and a line width
         of 10$\cm^{-1}$ is assigned.
         }
\end{figure*}

\begin{figure*}
\centering{
\includegraphics[scale=0.5,clip]{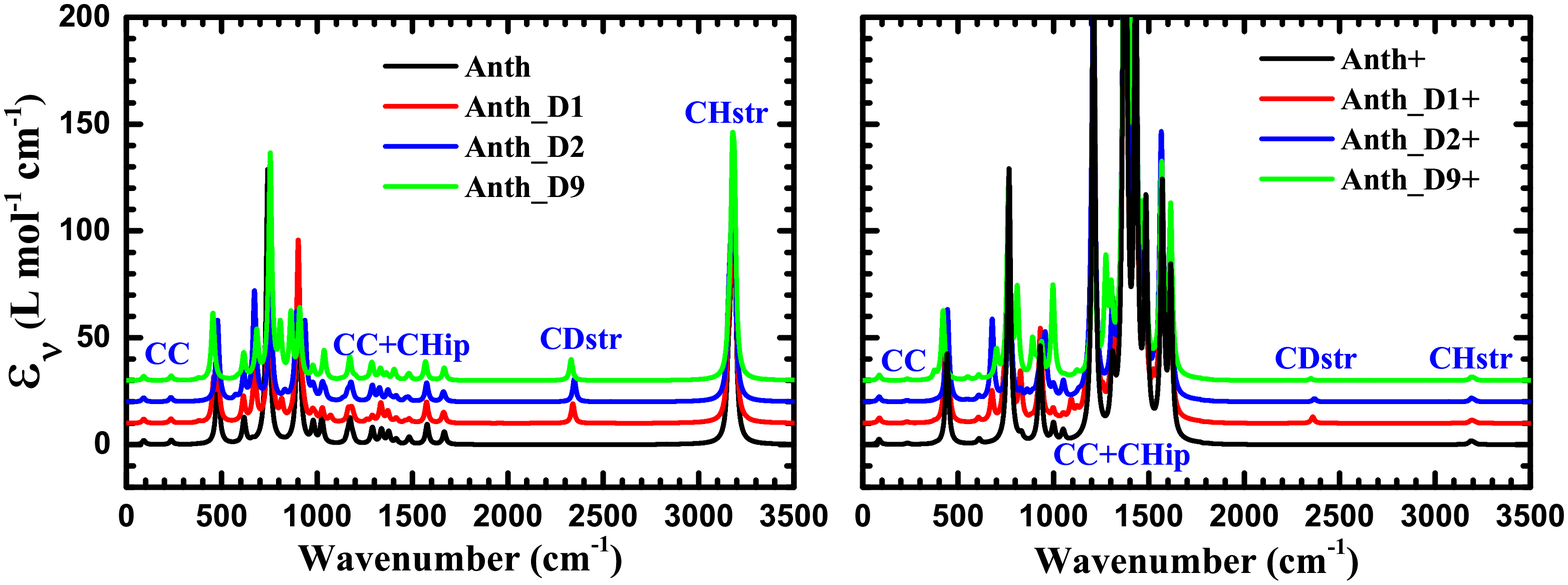}
}
\caption{\footnotesize
         \label{fig:Anth_IR_spec_all}
         Comparison of the computational spectra
         (red, blue and green lines)
         of mono-deuterated anthracene (left panel)
         and its cation (right panel)
         with that of pure anthracene and its cation (black lines).
         The frequencies are scaled with
         a factor of 0.9688, and a line width
         of 10$\cm^{-1}$ is assigned.
         }
\end{figure*}

\begin{figure*}
\centering{
\includegraphics[scale=0.5,clip]{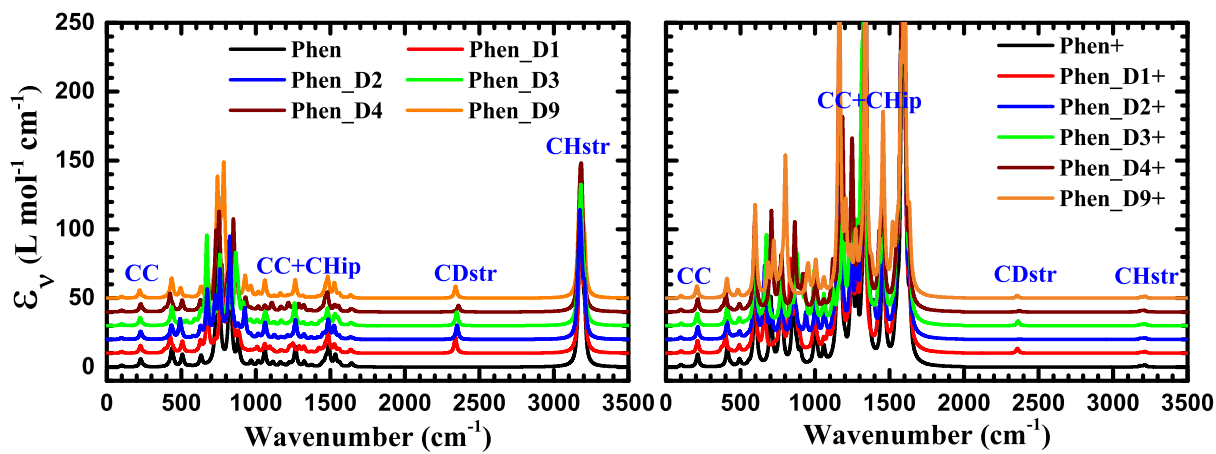}
}
\caption{\footnotesize
         \label{fig:Phen_IR_spec_all}
         Comparison of the computational spectra
         (red, blue, brown, green and yellow lines)
         of mono-deuterated phenanthrene (left panel)
         and its cation (right panel)
         with that of pure phenanthrene and its cation (black lines).
         The frequencies are scaled with
         a factor of 0.9688, and a line width
         of 10$\cm^{-1}$ is assigned.
         }
\end{figure*}

\begin{figure*}
\centering{
\includegraphics[scale=0.5,clip]{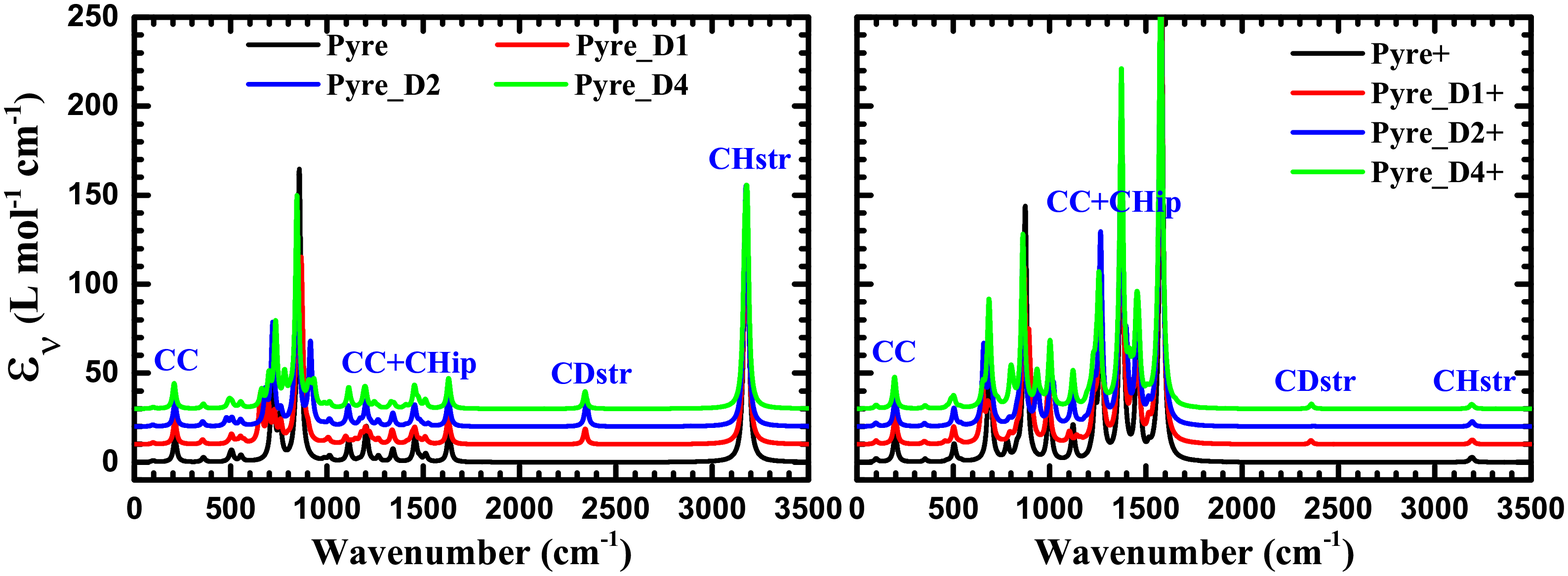}
}
\caption{\footnotesize
         \label{fig:Pyre_IR_spec_all}
         Comparison of the computational spectra
         (red, blue and green lines)
         of mono-deuterated pyrene (left panel)
         and its cation (right panel)
         with that of pure pyrene and its cation (black lines).
         The frequencies are scaled with
         a factor of 0.9688, and a line width
         of 10$\cm^{-1}$ is assigned.
         }
\end{figure*}

\begin{figure*}
\centering{
\includegraphics[scale=0.5,clip]{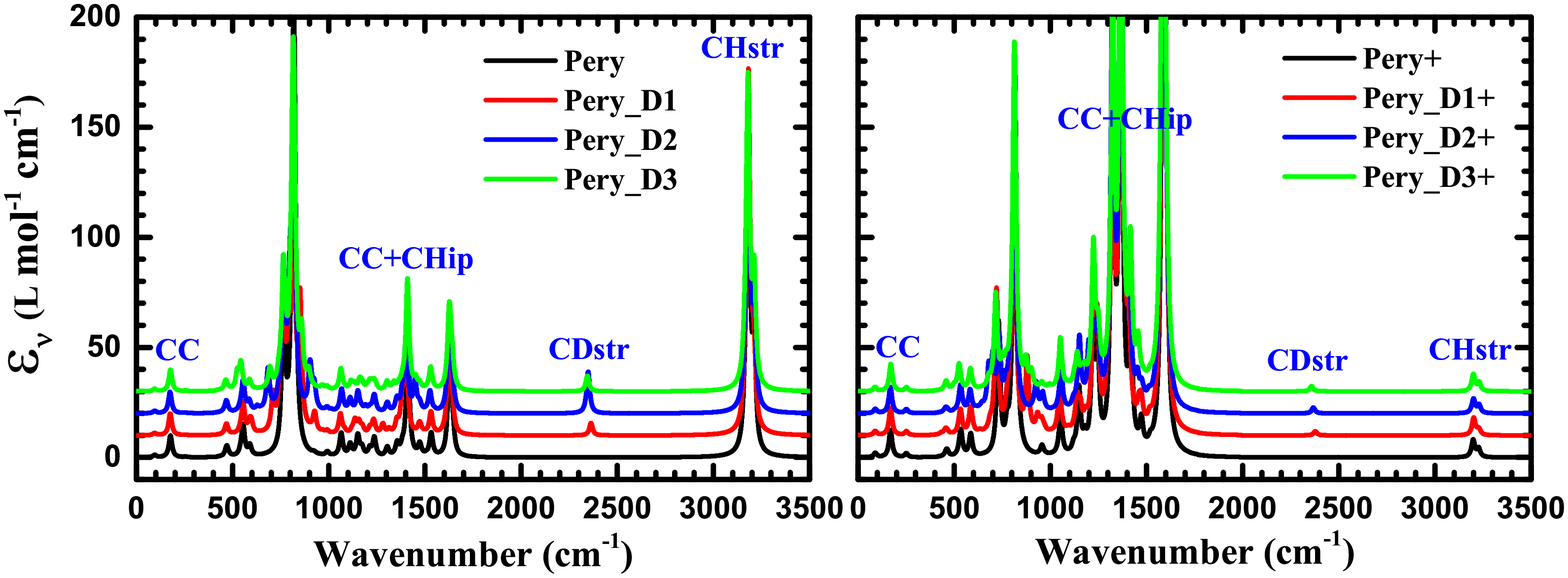}
}
\caption{\footnotesize
         \label{fig:Pery_IR_spec_all}
         Comparison of the computational spectra
         (red, blue and green lines)
         of mono-deuterated perylene (left panel)
         and its cation (right panel)
         with that of pure perylene and its cation (black lines).
         The frequencies are scaled with
         a factor of 0.9688, and a line width
         of 10$\cm^{-1}$ is assigned.
         }
\end{figure*}

\begin{figure*}
\centering{
\includegraphics[scale=0.5,clip]{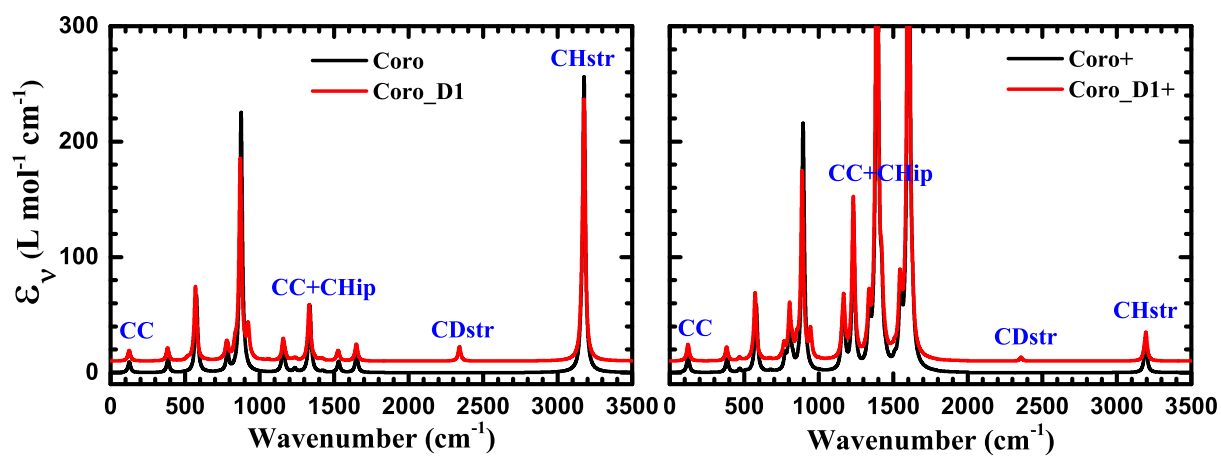}
}
\caption{\footnotesize
         \label{fig:Coro_IR_spec_all}
         Comparison of the computational spectra (red lines)
         of mono-deuterated coronene (left panel)
         and its cation (right panel)
         with that of pure coronene and its cation (black lines).
         The frequencies are scaled with
         a factor of 0.9688, and a line width
         of 10$\cm^{-1}$ is assigned.
         }
\end{figure*}

Considering the relatively low overall cosmic
D/H abundance (see \S\ref{sec:intro}),
it is reasonable to assume that singly-deuterated
PAHs play a more important role in astrophysics.
Meanwhile, we focus on the 3.3$\mum$ C--H stretch
and the 4.4$\mum$ C--D stretch, which mainly originate
from small PAH molecules of $\simali$20--30 C atoms
(see Figure~7 of Draine \& Li 2007).
Therefore, we calculate the IR spectra of
seven small mono-deuterated PAH species
and their cations:
$d$-benzene (C$_6$H$_5$D),
$d$-naphthalene (C$_{10}$H$_7$D),
$d$-anthracene (C$_{14}$H$_{9}$D),
$d$-phenanthrene (C$_{14}$H$_{9}$D),
$d$-pyrene (C$_{16}$H$_{9}$D),
$d$-perylene (C$_{20}$H$_{11}$D), and
$d$-coronene (C$_{24}$H$_{11}$D).
For each PAH species, we consider
all the mono-deuterated isomers
of which each has a different position
of D-atom attachment
(see Figure~\ref{fig:1D_PAH_structures}).
%
%
The calculated spectra are shown in
Figures~\ref{fig:Benz_IR_spec_all}--\ref{fig:Coro_IR_spec_all}.
The frequencies are scaled with a factor of 0.9688, and
the line widths are taken to be 10$\cm^{-1}$.\footnote{%
   Such a line width of 10$\cm^{-1}$ is adopted
   to distinctly reveal the major C--H, C--D and
   C--C vibrational bands in the computational spectra.
   Too large a line width would lead to the clumping of
   some of the major bands and therefore it becomes
   difficult to identify individual bands.
   We note that we assign a line width of
   20$\cm^{-1}$ for Benz$\_$6D
   (see \S\ref{subsec:PerdeuteratedPAH_Results},
    Figure~\ref{fig:PerDeu_Benzene}a
    and Figure~\ref{fig:PerDeu_spec_com}a)
   and 4$\cm^{-1}$ for Naph$\_$8D,
   Phen$\_$10D and Pyre$\_$10D
   (see \S\ref{subsec:PerdeuteratedPAH_Results},
    and Figure~\ref{fig:PerDeu_spec_com}a)
   since the computational spectra
   with these line widths best match
   the experimental spectra.
   }
To highlight the C--D features,
we also show in the corresponding panels
the IR spectra computed for pure PAHs.

As expected, the C--D stretching features are apparent
in the computed spectra at $\simali$4.4$\mum$
(i.e., 2270$\cm^{-1}$)
of all the deuterated derivatives.
The frequencies of the C--D stretch
for the deuterated cations are somewhat
blue-shifted with respect to the neutrals,
with an average wavelength
of $\simali$4.40$\mum$ for the neutrals
and $\simali$4.37$\mum$ for the cations.
The C--H stretch also shows a similar behavior
for both pure and deuterated PAHs,
with an average wavelength of $\simali$3.25$\mum$
for the neutrals and $\simali$3.22$\mum$ for the cations.
Meanwhile, the C--D and C--H stretches of the cations
are significantly suppressed with respect to the neutrals,
just like pure PAHs (e.g., see Allamandola et al.\ 1999),
superhydrogenated PAHs (Yang et al.\ 2020)
and PAHs with aliphatic functional groups (Yang et al.\ 2017).
Let $\Aaro$ and $\Acd$ respectively be the intrinsic band
strengths of the aromatic C--H and C--D stretches
on a per bond basis.
For the neutrals, the average $\Aaro$ and $\Acd$
are respectively $\simali$13.22$\km\mol^{-1}$
and $\simali$7.29$\km\mol^{-1}$,
while $\simali$0.63$\km\mol^{-1}$
and $\simali$1.38$\km\mol^{-1}$ for the cations.

The C--D oop bending ($\simali$15.4$\mum$) and
in-plane bending ($\simali$11.7$\mum$) features
are well mixed with the C--C--C or C--H related features,
e.g, the latter often overlaps with the C--H oop bending
features (see Hudgins et al.\ 2004).
Therefore, it is difficult to identify the C--D in-plane and
oop bending bands from the computed IR spectra
since these features lie so close to the C--H and
C--C--C bands that no matter how small a line width
is assigned these features always overlap.
Nevertheless, we can still identify them
from the Gaussian09 output files and
read their frequencies and intensities,
as tabulated in Table~\ref{tab:Freq_Int_DPAH_all}
and Table~\ref{tab:Freq_Int_DPAHPlus_all}.
For both neutrals and cations,
the frequencies are basically in the wavenumber range
of $\simali$830--960$\cm^{-1}$
for the C--D in-plane bending modes
and $\simali$600--675$\cm^{-1}$
for the C--D oop bending modes.

For both neutrals and cations,
the band strengths of these two vibrational modes
scatter significantly from one molecule to another,
with the standard deviations at the same order of
magnitude as the average values
(see Table~\ref{tab:Freq_Int_DPAH_all}
and Table~\ref{tab:Freq_Int_DPAHPlus_all}).
Moreover, the vibrations of the C--D oop bending
and in-plane bending are usually coupled with other
vibrational modes. For these reasons, we will not discuss
any further the intensities of the C--D bending modes.

\section{Astrophysics Implications}\label{sec:discussion}
\subsection{Aromatic C--H and C--D Band Intensities of Deuterated PAHs}\label{sec:BandStrengths}
\begin{figure*}
\centering{
\includegraphics[scale=0.6,clip]{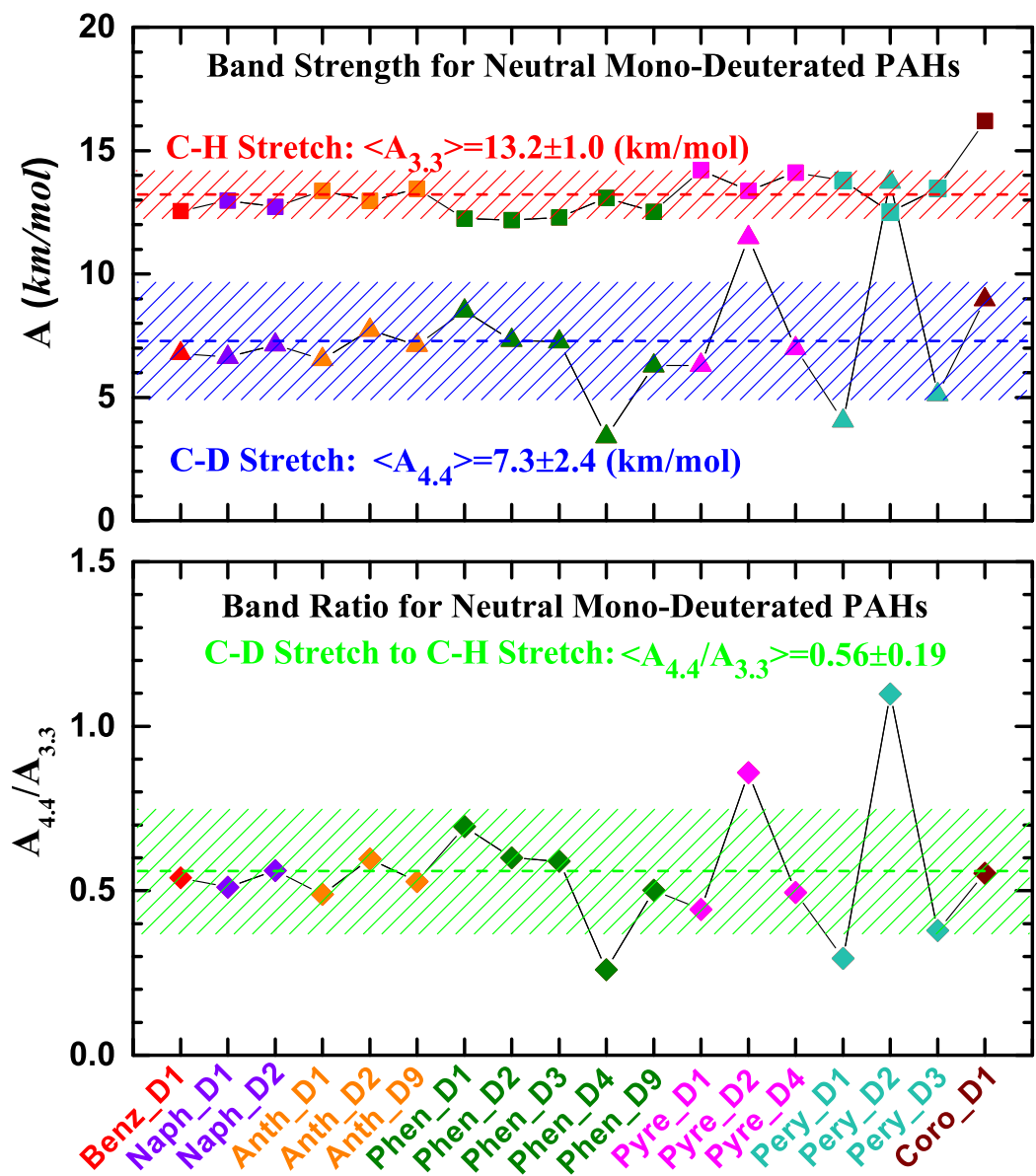}
}
\caption{\footnotesize
         \label{fig:PAH_1D_A44A33_all}
         Band strengths of the 3.3$\mum$ C--H stretches ($\ACH$)
         and the 4.4$\mum$ C--D stretches ($\ACD$)
         as well as the band-strength ratios $\ACD/\ACH$
         computed at level {\rm B3LYP/6-311+G$^{\ast\ast}$}
         for the mono-deuterated {\it neutral} PAH molecules
         shown in Figure~\ref{fig:1D_PAH_structures}.
         }
\end{figure*}

We show in Figure~\ref{fig:PAH_1D_A44A33_all}
the intensities of the aromatic C--H and C--D stretches
computed for 18 mono-deuterated {\it neutral} PAH species.
These species are displayed in the order of
their (increasing) sizes from benzene,
naphthalene, ..., up to coronene.
For the same parental molecule,
the isomers are displayed in the order of
the position where the D atom is attached
(e.g., from Phen$\_$D1, Phen$\_$D2, ...,
to Phen$\_$D9).
Except for one species (i.e., Coro$_{-}$D1),
the band strength of the C--H stretch at 3.3$\mum$
on a per unit C--H bond basis,
does not vary much among different molecules.
On average, these mono-deuterated neutral PAH species
(including Coro$_{-}$D1)  have a mean (per-bond) band strength
of $\langle\Aaro\rangle\approx13.2\pm1.0\km\mol^{-1}$.
The band strength of the C--D stretch at 4.4$\mum$,
on a per unit C--D bond basis, $\ACD$,
exhibits more appreciable variations than $\Aaro$
among these 18 species, with three species
(i.e., Phen$\_$D4, Pyre$\_$D2 and Pery$\_$D2)
noticeably deviating from the majority.
On average, these mono-deuterated neutral species
have a mean band strength
of $\langle\ACD\rangle\approx7.3\pm2.4\km\mol^{-1}$.
For these mono-deuterated neutral PAHs,
the mean band-strength ratio of the C--D stretch
to the C--H stretch is
$\langle\ACD/\Aaro\rangle\approx0.56\pm0.19$.

\begin{figure*}
\centering{
\includegraphics[scale=0.5,clip]{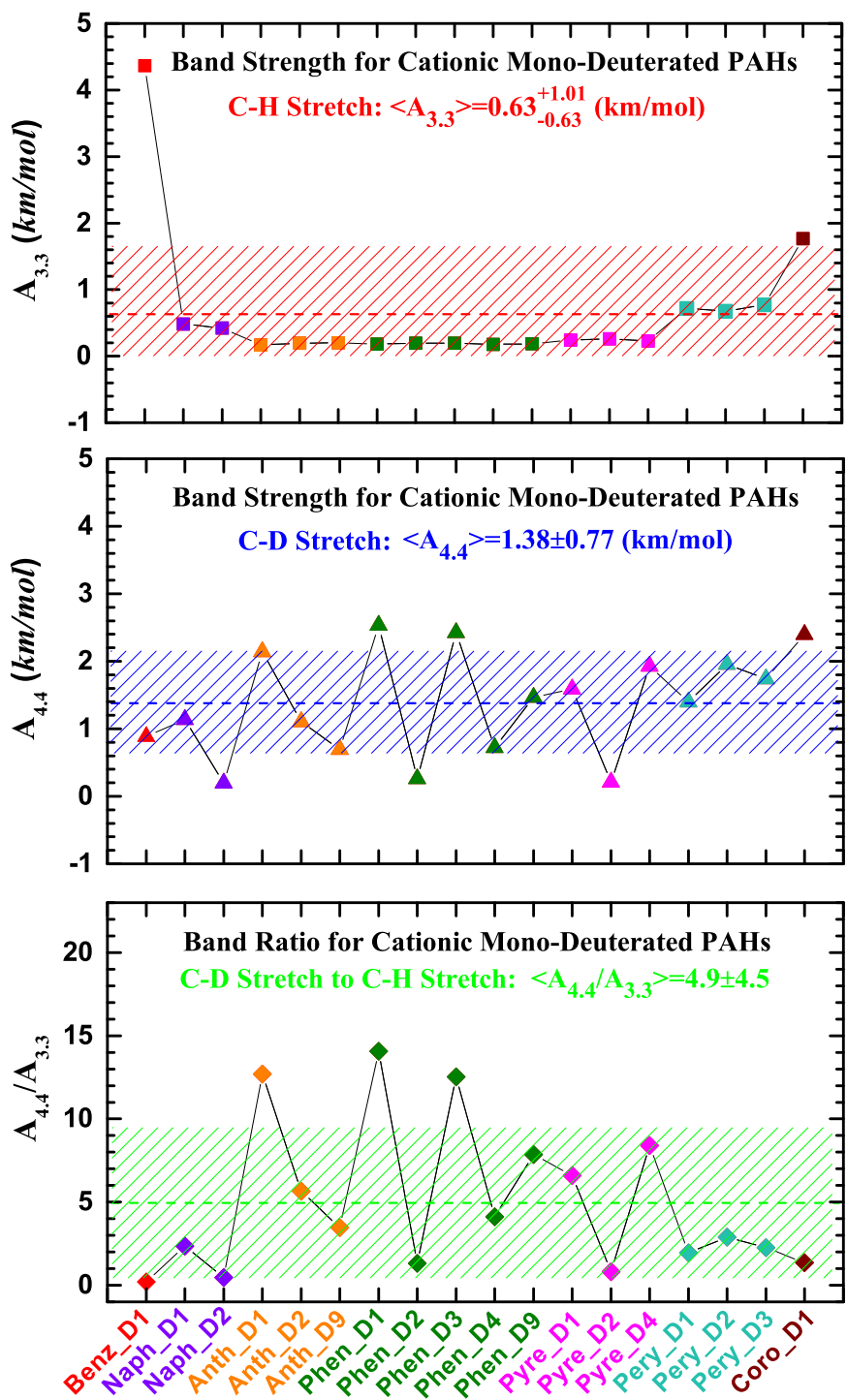}
}
\caption{\footnotesize
         \label{fig:PAH_1DPlus_A44A33_all}
         Band strengths of the 3.3$\mum$ C--H stretches ($\ACH$)
         and the 4.4$\mum$ C--D stretches ($\ACD$)
         as well as the band-strength ratios $\ACD/\ACH$
         computed at level {\rm B3LYP/6-311+G$^{\ast\ast}$}
         for the cationic counterparts of the mono-deuterated PAHs
         shown in Figure~\ref{fig:1D_PAH_structures}.
         }
\end{figure*}

Similarly, we show in Figure~\ref{fig:PAH_1DPlus_A44A33_all}
the band strengths of the C--H and C--D stretches
computed for 18 mono-deuterated {\it cationic} PAH species.
Except for one species (i.e., Benz$_{-}$D1),
$\Aaro$ varies very little among different molecules.
In contrast, $\ACD$ exhibits more significant variations.
With $\langle\Aaro\rangle\approx0.63^{+1.01}_{-0.63}\km\mol^{-1}$
and $\langle\ACD\rangle\approx1.38\pm0.77\km\mol^{-1}$,
it is apparent that cationic PAHs have their C--H and C--D stretches
substantially suppressed compared to their neutral counterparts.
On average, for these mono-deuterated cationic PAHs
the mean C--D to C--H band-strength ratio is
$\langle\ACD/\Aaro\rangle\approx4.9\pm4.5$.

For mono-deuterated neutral PAHs,
at first glance, it appears that large,
compact pericondensed PAHs intend
to exhibit more appreciable variations
in the C--D stretching band-strength $\ACD$
(see Figure~\ref{fig:PAH_1D_A44A33_all}).
However, a close inspection of
Figure~\ref{fig:PAH_1DPlus_A44A33_all}
reveals that,
for mono-deuterated PAH cations,
small, catacondensed PAHs also exhibit
appreciable variations in $\ACD$.
Apparently, there does not seem to
exist a simple dependence of $\ACD$
on the PAH size and structure.

In Table~\ref{tab:MeanBandStrengths}
we summarize the {\it mean} band strengths
of the C--H and C--D stretches
computed for mono-deuterated PAHs,
obtained by averaging over
all the mono-deuterated species,
with recommended values of
$\langle\Aaro\rangle\approx13.2\pm1.0\km\mol^{-1}$,
$\langle\ACD\rangle\approx7.3\pm2.4\km\mol^{-1}$
and $\langle\ACD/\Aaro\rangle\approx0.56\pm0.19$
for neutral PAHs, and
$\langle\Aaro\rangle\approx0.63^{+1.01}_{-0.63}\km\mol^{-1}$,
$\langle\ACD\rangle\approx1.38\pm0.77\km\mol^{-1}$
and $\langle\ACD/\Aaro\rangle\approx4.94\pm4.47$
for cations.

\subsection{Deuterium Depletion onto PAHs}
Let $\Dpah$ be the degree of deuteration of PAHs.
For a deuterated PAH molecule containing
$\NC$ aromatic C atoms, $\NH$ H atoms,
and $\ND$ D atoms, we define
the degree of deuteration of the molecule
as $\Dpah\equiv\ND/\left(\NH+\ND\right)$.
Let $\Iratioobs$ be the {\it observed} ratio
of the power emitted from the 4.4$\mum$
aromatic C--D feature ($I_{4.4}$) to that from
the 3.3$\mum$ aromatic C--H feature ($I_{3.3}$),
we have
\begin{equation}\label{eq:Iratio}
\left(\frac{I_{4.4}}{I_{3.3}}\right)_{\rm obs}
\approx \left(\frac{A_{4.4}}{A_{3.3}}\right)
\times\left(\frac{\NCD}{\NCH}\right)
\times\left(\frac{B_{4.4}}{B_{3.3}}\right) ~~,
\end{equation}
where $\NCH$ and $\NCD$ are respectively
the number of C--H and C--D bonds in
a deuterated PAH molecule,
$B_\lambda(T)$ is the Planck function
at wavelength $\lambda$ and temperature $T$,
and $A_{3.3}$ and $A_{4.4}$
are the intrinsic band strengths
of the aromatic C--H and C--D stretches
(on a per C--H or C--D bond basis).
Since, as discussed earlier
(see \S\ref{subsec:MonoD_Results} 
and \S\ref{sec:BandStrengths}),
the C--H and C--D stretching features
are predominantly emitted by neutral PAHs,
we will adopt $\ACD/\ACH\approx0.56\pm0.19$
(see  \S\ref{sec:BandStrengths}
and Table~\ref{tab:MeanBandStrengths}).

Assuming in a PAH molecule all the D and
H atoms are attached to aromatic C atoms,
then $\NCD=\ND$ and $\NCH=\NH$,\footnote{%
    We note that interstellar PAHs could be superhydrogenated
    (Bernstein et al.\ 1996, Sandford et al.\ 2013, Yang et al.\ 2020)
    or attached with aliphatic functional groups
    (Geballe et al.\ 1985, Sandford 1991,
     Kwok \& Zhang 2011, Yang et al.\ 2017)
    which could convert the aromatic C atoms to aliphatic.
    However, we have shown that the degree of
    superhydrogenation (Yang et al.\ 2020)
    and the aliphatic content of PAHs are minor
    (Li \& Draine 2012, Yang et al.\ 2013, 2017).
    Nevertheless, the aliphatic C--D stretches
    at 4.65 and 4.75$\mum$ appear to be present
    in the ISO/SWS and AKARI/IRC spectra
    of several interstellar objects
    (Peeters et al.\ 2004, Onaka et al.\ 2014,
    Doney et al.\ 2016),
    indicating that some D atoms may be attached
    to aliphatic C atoms.
    }
the degree of deuteration of the molecule becomes
\begin{equation}
\label{eq:NDNH}
 \Dpah \approx
\left\{
1\,+\,\left(\frac{I_{3.3}}{I_{4.4}}\right)_{\rm obs}
\times\left(\frac{\Acd}{\Aaro}\right)
\times \left(\frac{B_{4.4}}{B_{3.3}}\right)
\right\}^{-1} ~~.
\end{equation}
%
%
%
The 3.3$\mum$ C--H stretch
and 4.4$\mum$ C--D stretch
are most effectively emitted by PAHs
of vibrational temperatures of
$\simali$728$\K$ and $\simali$546$\K$,
respectively, stochastically heated by individual
ultraviolet photons (Draine \& Li 2001).\footnote{%
  Let $C_{\rm abs}(\lambda)$ be the dust
  absorption cross section at wavelength $\lambda$
  and $j_\lambda$ be the dust IR emissivity.
  For $C_{\rm abs}(\lambda)\propto \lambda^{-\beta}$,
  $\lambda j_\lambda$ peaks at
  $\lambda_p\approx\left(h c/k T\right)/\left(4+\beta\right)$,
  where $h$ is the Planck constant,
  $c$ is the speed of light, and $k$ is
  the Boltzmann constant (see Li 2009).
  With $\beta\approx2$ for PAH-like molecules,
  we have $T\approx728\K$ for $\lambda_p\sim3.3\mum$
  and $T\approx546\K$ for $\lambda_p\sim4.4\mum$.
  }
For $546\simlt T\simlt 728\K$,
we obtain $B_{3.3}/B_{4.4}\approx0.75\pm0.10$.
By relaxing the temperature range to
$400\simlt T\simlt 900\K$,
we obtain $B_{3.3}/B_{4.4}\approx0.70\pm0.28$.\footnote{%
   Even if we extend the temperature range to
   $300\simlt T\simlt 1000\K$,
   $B_{3.3}/B_{4.4}\approx0.63\pm0.38$
   is only lower by $\simali$10\%.
   }
In the following, we will adopt
$B_{3.3}/B_{4.4}\approx0.70\pm0.28$
(suitable for $400\simlt T\simlt 900\K$).

Many observational efforts have been made
to detect deuterated PAHs through their
vibrational C--D bands.
Peeters et al.\ (2004) searched for
the C--D emission feature of deuterated PAHs
in the Orion Bar and M17 photodissociation regions (PDRs).
Based on ISO/SWS observations,
Peeters et al.\ (2004) reported detection of
weak emission features at 4.4$\mum$
and 4.65$\mum$ in the Orion Bar.
In M17, they confirmed the earlier detection
of the weak emission feature at 4.65$\mum$ reported
by Verstraete et al.\ (1996). Peeters et al.\ (2004)
attributed the 4.4$\mum$ feature to the aromatic C--D
stretch in deuterated PAHs and the 4.65$\mum$ feature
to the aliphatic C--D stretch in super-deuterated
PAHs or PAHs with deuterated aliphatic side groups.
Onaka et al.\ (2014) have also searched for
the C--D emission feature of deuterated PAHs
in the Orion Bar and M17 as well as in G18.14.0,
a reflection nebula, based on the near-IR spectroscopy
at $\simali$2.5--5$\mum$ obtained by AKARI/IRC.
%
Also with AKARI/IRC, Doney et al.\ (2016) have performed
a deuterated PAH survey in 53 H{\sc ii} regions
in the Milky Way, Large Magellanic Cloud
and Small Magellanic Cloud.
They reported the detection of
deuterated PAHs in six sources,
as revealed by, in addition to
the aromatic C--D stretching emission band
at 4.4$\mum$, the asymmetric aliphatic
C--D stretch at $\simali$4.63$\mum$
and the symmetric aliphatic C--D stretch
at $\simali$4.75$\mum$.

We have compiled all the available observational data
on $\left(I_{4.4}/I_{3.3}\right)_{\rm obs}$
(Peeters et al.\ 2004, Onaka et al.\ 2014, Doney et al.\ 2016)
and summarized these data in Table~\ref{tab:I44I33_obs}.
On average, we obtain
$\langle\Iratioobs\rangle\approx0.019$.
With $B_{3.3}/B_{4.4}\approx0.70\pm0.28$,
we obtain from $\Iratioobs\approx0.019$
a deuteration degree
of $\Dpah\approx2.4\%$
for neutral PAHs
of $\Acd/\Aaro\approx0.56\pm0.16$.
%
Compared to the interstellar abundance
of $\Dism\approx2\times10^{-5}$,
this implies a D-enrichment in PAHs
by a factor of $\Dpah/\Dism\approx1200$.
%
%

We note that the actual degree of deuteration
in PAHs could be higher than 2.4\%
since some D atoms may be incorporated
into PAHs to generate aliphatic C--D bands,
either in the form of excess D atoms attached
to fully hydrogenated or superhydrogenated PAHs
(Hudgins et al.\ 2004;
Buragohain et al.\ 2015, 2016)
or in the form of D-substitued alphatic sidegroups
(Buragohain et al.\ 2020).
Both experimental measurements and
DFT-based computations have demonstrated
that the IR spectra of such D-containing PAHs
exhibit a prominent aliphatic C--D stretching
band at $\simali$4.6--4.8$\mum$
(Hudgins et al.\ 2004;
Buragohain et al.\ 2015, 2016, 2020).
%
%
Indeed, ISO/SWS and AKARI/IRC observations
have shown that whenever the 4.4$\mum$
aromatic C--D band was detected,
an accompanying aliphatic C--D band
was also detected. The latter is often stronger
than the former by a factor of $\simali$2--7
(Peeters et al.\ 2004, Doney et al.\ 2006).
%
This suggests that a significant amount of D atoms
may be attached to aliphatic C atoms.\footnote{%
    It is not clear whether or to what extent
    the strength and frequency of the aromatic 
    C--D stretch would be affected when a deuterated
    PAH molecule is also attached by D-substituted
    aliphatic sidegroups (e.g., --CH$_2$D) or some
    of its peripheral C atom have one H atom and 
    one extra D atom. Deuterated PAHs with D-substituted
    aliphatic sidegroups or excess D atoms are expected
    to exhibit both aromatic and aliphatic C--D stretches
    as seen in the ISO/SWS and AKARI spectra of 
    the Orion Bar, M17, and H{\sc ii} regions
    (Peeters et al.\ 2004, Doney et al.\ 2016).
    It would be interesting to experimentally and 
    computationally explore the frequencies and strengths
    of the aromatic and aliphatic C--H and C--D bands of
    such molecules. On the other hand, the aromatic C--D 
    band at 4.4$\mum$ and the aliphatic C--D band
    at $\simali$4.65--4.8$\mum$ 
    seen in the Orion Bar, M17, and H{\sc ii} regions 
    may not necessarily be emitted by the same molecule
    (i.e., the former could be emitted by deuterated PAHs
     while the latter is emitted by PAHs attached 
     by D-substituted aliphatic sidegroups 
     or excess D atoms).
    }
Also, the current estimation is based on
the 3.3 and 4.4$\mum$ features,
which are only sensitive to small PAHs.
If an appreciable amount of D atoms
reside in large PAHs, they will emit mainly
at longer wavelengths through C--D oop
and in-plane bending modes.
Future observational studies with JWST of
both aliphatic and aromatic C--D stretching bands
and aromatic C--D oop and in-plane bending bands
will be highly valuable.

\section{Summary}\label{sec:summary}
To facilitate a quantitative understanding
of PAHs as a possible reservoir of interstellar D,
we have employed the hybrid DFT method B3LYP
in conjunction with the 6-311+G$^{\ast\ast}$ basis set
to calculate the IR vibrational spectra of
singly deuterated PAHs and their cations
of various sizes (from benzene, naphthalene
to perylene and coronene).
The major results are as follows:
\begin{enumerate}
\item The aromatic C--D stretching, in-plane bending
          and oop bending bands are seen at $\simali$4.4$\mum$,
         $\simali$11.7$\mum$ and $\simali$15.4$\mum$,
         respectively, in all the deuterated PAH species.
\item For all these mono-deuterated species,
          we have derived from the computed spectra
          the intrinsic band strengths of
          the 3.3$\mum$ aromatic C--H stretch ($\Aaro$)
          and the 4.4$\mum$ aromatic C--D stretch ($\Acd$).
          Both the C--H and C--D stretches predominantly
          arise from neutral PAHs.
          By averaging over all these molecules,
          we have determined the mean band strengths to be
          $\langle\Aaro\rangle\approx13.2\pm1.0\km\mol^{-1}$
          and $\langle\ACD\rangle\approx7.3\pm2.4\km\mol^{-1}$
          for neutral PAHs, and
          $\langle\Aaro\rangle\approx0.63^{+1.01}_{-0.63}\km\mol^{-1}$,
          $\langle\ACD\rangle\approx1.38\pm0.77\km\mol^{-1}$
          for cationic PAHs. The mean band-strength ratios of
          the C--D stretch to the C--H stretch are
          $\langle \Acd/\Aaro\rangle\approx 0.56\pm0.19$ for the neutrals,
          and $\langle\Acd/\Aaro\rangle\approx 4.94\pm4.47$ for the cations.
\item We have also derived from
          the mono-deuterated species
          the intrinsic strengths of
          the 11.7$\mum$ aromatic C--D
          in-plane bending band ($A_{11.7}$) and
          the 15.4$\mum$ aromatic C--D
          oop bending band ($A_{15.4}$).
          By averaging over all these mono-deuterated species,
          we have determined the mean band strengths to be
          $\langle A_{11.7}\rangle\approx 2.65\km\mol^{-1}$ and
          $\langle A_{15.4}\rangle\approx 19.8\km\mol^{-1}$
          for the neutrals, and
          $\langle A_{11.7}\rangle\approx 5.56\km\mol^{-1}$ and
          $\langle A_{15.4}\rangle\approx 18.86\km\mol^{-1}$
          for the cations.
\item We have compiled all the observational data
          from ISO/SWS and AKARI/IRC and derived
          the ratio of the power emitted from
          the astronomical 4.4$\mum$
          aromatic C--D feature ($I_{4.4}$) to that from
          the astronomical 3.3$\mum$
          aromatic C--H feature ($I_{3.3}$) to be
          $\langle\left(I_{4.4}/I_{3.3}\right)_{\rm obs}\rangle\approx0.019$.
          By comparing the computationally-derived mean ratio
          of $\langle\Acd/\Aaro\rangle\approx 0.56$ for
          neutral deuterated PAHs, we have estimated
          the degree of deuteration to be $\Dpah\approx2.4\%$
          for neutral PAHs which dominate the emission
          at the C--H and C--D stretching bands.
          Compared to the interstellar abundance
          of $\Dism\approx2\times10^{-5}$,
          this implies that interstellar PAHs are
          D-enriched by a factor of
          $\Dpah/\Dism\approx1200$.
          %
          %
%
\end{enumerate}

\acknowledgments{%
We thank B.T.~Draine and the anonymous referees
for very helpful suggestions.
XJY is supported in part by NSFC 11873041, 11473023
and the NSFC-CAS Joint Research Funds
in Astronomy (U1731106, U1731107).
AL is supported in part by NASA grants
80NSSC19K0572 and 80NSSC19K0701.
RG is supported in part by NSF-PRISM grant
Mathematics and Life Sciences (0928053).
Computations were performed using the high-performance computer
resources of the University of Missouri Bioinformatics Consortium.
}



\clearpage

\begin{table*}
\footnotesize
\begin{center}
\caption[]{\footnotesize
           Comparison of the Experimental and Computational
           Frequencies and Relative Intensities
           of the C--D Bands for Perdeuterated PAHs.
           }
\label{tab:Freq_Int_Comp_CalExp}
\begin{tabular}{lcccccc}
\noalign{\smallskip} \hline \hline \noalign{\smallskip}
\multicolumn{1}{c}{  }&	\multicolumn{1}{c}{  }  &    \multicolumn{2}{c}{ Experimental } & \multicolumn{3}{c}{ Calculational } \\\noalign{\smallskip}\cline{3-4}\cline{5-7}\noalign{\smallskip}
Vibrational Mode   &  Molecule   &
\multicolumn{1}{c}{$\tilde\nu^\dagger$}   &  \multicolumn{1}{c}{$A_{\rm
    CD}/A_{\rm CD,oop}$}  &  \multicolumn{1}{c}{$\tilde{\nu}$$\dagger$}   &
\multicolumn{1}{c}{$A_{\rm CD}^\ddagger$} & \multicolumn{1}{c}{$A_{\rm
    CD}/A_{\rm CD,oop}$} \\\noalign{\smallskip}\hline
\multirow{3}{*}{C--D OOP Bending}           &   Naph$\_$8D  &  	630.9, 634.3	&	1.00	&	620.6	        &	6.16   &1.00    \\
                                            &   Phen$\_$10D &	610.7	        &	1.00	&	601.8, 638.5	&	4.23   &1.00    \\
                                            &   Pyre$\_$10D &	604.0	        &	1.00	&	593.1	        &	5.09   &1.00    \\\hline
\multirow{3}{*}{C--D In-Plane Bending}      &   Naph$\_$8D  & 	829.8, 832.5	&	0.17	&	816.6, 824.3	&	0.97   &0.16 	\\
                                            &   Phen$\_$10D &	834.5   	    &	0.44 	&	821.1, 830.9	&	0.91   &0.21 	\\
                                            &   Pyre$\_$10D &	841.3	        &	0.18	&	822.3, 828.3	&	0.96   &0.19    \\\hline
\multirow{3}{*}{C--D Stretching }           &   Naph$\_$8D  &	2250$\sim$2315	&	1.12	&	2255$\sim$2288	&	6.91   &1.12 	\\
                                            &   Phen$\_$10D &	2249$\sim$2311	&	1.42	&	2253$\sim$2297	&	6.61   &1.56 	\\
                                            &   Pyre$\_$10D &	2245$\sim$2313	&	1.51	&	2253$\sim$2285	&	7.64   &1.50 	\\
\hline
\noalign{\smallskip} \noalign{\smallskip}
\end{tabular}

\begin{description}
\item[$^\dagger$] Vibrational frequencies or wavenumbers in $\cm^{-1}$
\item[$^\ddagger$] Vibrational intensities (per C--D bond) in $\km\mol^{-1}$
\end{description}

\end{center}
\end{table*}

\begin{table*}
\footnotesize
\begin{center}
\caption[]{\footnotesize
           Wavelengths ($\lambda$) and Intensities
           ($A_\lambda$ in $\km\mol^{-1}$)
           of the Nominal ``3.3$\mum$'' C--H Stretching,
           ``4.4$\mum$'' C--D Stretching,
           ``11.7$\mum$'' C--D In-Plane Bending, and
           ``15.4$\mum$'' C--D OOP Bending Bands
           Computed at the B3LYP/6-311+G$^{\ast\ast}$ Level
           for All the Neutral Mono-Deuterated PAHs
           Shown in Figure~\ref{fig:1D_PAH_structures}.
           The Intensities $A_\lambda$ Relates to the Molar
           Extinction Coefficients $\varepsilonWV$ through
           Eq.\,\ref{eq:epsilon2A}.
           }
\label{tab:Freq_Int_DPAH_all}
\begin{tabular}{lccccccccr}
\noalign{\smallskip} \hline \hline \noalign{\smallskip}
Compound	    &	$\lambda_{3.3}$ & $A_{3.3}$
                &	$\lambda_{4.4}$ & $A_{4.4}$
                &   $\Aratio$
                &	$\lambda_{11.7}$  & $A_{11.7}$
                &	$\lambda_{15.4}$  & $A_{15.4}$          \\
                &  ($\mu$m)  &  (km$\mol^{-1}$)
                &  ($\mu$m)  & (km$\mol^{-1}$)  &
                &  ($\mu$m)  &  (km$\mol^{-1}$)
                &  ($\mu$m)  & (km$\mol^{-1}$)  \\
\noalign{\smallskip} \hline \noalign{\smallskip}
Benz$\_$D1	&	3.24 	&	12.55 	&	4.40 	&	6.78 	&	0.54 	&	11.84 	&	2.91 	&	16.76 	&	71.34 	\\
Naph$\_$D1	&	3.25 	&	12.98 	&	4.41 	&	6.62 	&	0.51 	&	11.67 	&	2.77 	&	15.31 	&	1.15 	\\
Naph$\_$D2	&	3.25 	&	12.71 	&	4.40 	&	7.13 	&	0.56 	&	11.48 	&	1.68 	&	15.28 	&	8.50 	\\
Anth$\_$D1	&	3.25 	&	13.37 	&	4.41 	&	6.53 	&	0.49 	&	11.70 	&	2.03 	&	15.35 	&	18.72 	\\
Anth$\_$D2	&	3.25 	&	12.96 	&	4.39 	&	7.73 	&	0.60 	&	11.66 	&	2.07 	&	15.37 	&	36.57 	\\
Anth$\_$D9	&	3.25 	&	13.45 	&	4.43 	&	7.11 	&	0.53 	&	10.52 	&	0.41 	&	15.10 	&	15.40 	\\
Phen$\_$D1	&	3.24 	&	12.25 	&	4.41 	&	8.51 	&	0.69 	&	11.55 	&	2.97 	&	15.29 	&	25.58 	\\
Phen$\_$D2	&	3.24 	&	12.19 	&	4.39 	&	7.31 	&	0.60 	&	11.46 	&	1.78 	&	15.32 	&	24.78 	\\
Phen$\_$D3	&	3.24 	&	12.29 	&	4.40 	&	7.25 	&	0.59 	&	11.33 	&	2.76 	&	15.39 	&	46.35 	\\
Phen$\_$D4	&	3.24 	&	13.08 	&	4.37 	&	3.40 	&	0.26 	&	11.09 	&	6.24 	&	15.19 	&	10.11 	\\
Phen$\_$D9	&	3.24 	&	12.52 	&	4.41 	&	6.28 	&	0.50 	&	10.72 	&	2.63 	&	15.41 	&	0.21 	\\
Pyre$\_$D1	&	3.25 	&	14.21 	&	4.41 	&	6.30 	&	0.44 	&	11.45 	&	3.48 	&	15.76 	&	19.46 	\\
Pyre$\_$D2	&	3.25 	&	13.37 	&	4.40 	&	11.48 	&	0.86 	&	11.54 	&	0.53 	&	15.17 	&	12.85 	\\
Pyre$\_$D4	&	3.25 	&	14.10 	&	4.41 	&	6.97 	&	0.49 	&	11.12 	&	2.16 	&	15.62 	&	6.53 	\\
Pery$\_$D1	&	3.24 	&	13.78 	&	4.37 	&	4.05 	&	0.29 	&	11.13 	&	5.61 	&	--  	&	--  	\\
Pery$\_$D2	&	3.24 	&	12.50 	&	4.39 	&	13.72 	&	1.10 	&	11.44 	&	0.79 	&	15.09 	&	13.46 	\\
Pery$\_$D3	&	3.24 	&	13.47 	&	4.40 	&	5.11 	&	0.38 	&	11.52 	&	4.72 	&	14.88 	&	6.31 	\\
Coro$\_$D1	&	3.25 	&	16.20 	&	4.41 	&	8.96 	&	0.55 	&	11.45 	&	2.13 	&	--  	&	--  	\\\hline \noalign{\smallskip}
Average	&	3.25 	&	13.22 	&	4.40 	&	7.29 	&	0.56 	&	11.37 	&	2.65 	&	15.39 	&	19.83 	\\
Stdev	&	0.00 	&	0.97 	&	0.01 	&	2.39 	&	0.19 	&	0.34 	&	1.59 	&	0.42 	&	18.48 \\
\hline
\noalign{\smallskip} \noalign{\smallskip}
\end{tabular}
\end{center}
\end{table*}

\begin{table*}
\footnotesize
\begin{center}
\caption[]{\footnotesize
                 Same as Table~\ref{tab:Freq_Int_DPAH_all}
                 for Mono-Deuterated Cationic PAHs.
                 }
\label{tab:Freq_Int_DPAHPlus_all}
\begin{tabular}{lccccccccr}
\noalign{\smallskip} \hline \hline \noalign{\smallskip}
Compound	    &	$\lambda_{3.3}$ & $A_{3.3}$
                &	$\lambda_{4.4}$ & $A_{4.4}$
                &   $\Aratio$
                &	$\lambda_{11.7}$  & $A_{11.7}$
                &	$\lambda_{15.4}$  & $A_{15.4}$          \\
                &  ($\mu$m)  &  (km$\mol^{-1}$)
                &  ($\mu$m)  & (km$\mol^{-1}$)  &
                &  ($\mu$m)  &  (km$\mol^{-1}$)
                &  ($\mu$m)  & (km$\mol^{-1}$)  \\
\noalign{\smallskip} \hline \noalign{\smallskip}
Benz$\_$D1+	&	3.22	&	4.37	&	4.35	&	0.89	&	0.20 	&	12.22 	&	4.11 	&	16.73 	&	59.30 	\\
Naph$\_$D1+	&	3.22	&	0.48	&	4.37	&	1.14	&	2.36 	&	11.64 	&	1.00 	&	15.06 	&	3.31 	\\
Naph$\_$D2+	&	3.22	&	0.42	&	4.35	&	0.20	&	0.47 	&	11.42 	&	1.18 	&	15.40 	&	22.61 	\\
Anth$\_$D1+	&	3.23	&	0.17	&	4.37	&	2.14	&	12.69 	&	11.65 	&	0.00 	&	15.25 	&	10.03 	\\
Anth$\_$D2+	&	3.23	&	0.20	&	4.36	&	1.10	&	5.65 	&	12.02 	&	1.12 	&	15.74 	&	27.35 	\\
Anth$\_$D9+	&	3.23	&	0.20	&	4.39	&	0.69	&	3.47 	&	10.35 	&	27.75 	&	14.73 	&	9.01 	\\
Phen$\_$D1+	&	3.22	&	0.18	&	4.38	&	2.54	&	14.06 	&	11.50 	&	1.63 	&	15.61 	&	27.58 	\\
Phen$\_$D2+	&	3.22	&	0.20	&	4.35	&	0.26	&	1.32 	&	11.43 	&	2.07 	&	15.58 	&	36.72 	\\
Phen$\_$D3+	&	3.22	&	0.19	&	4.37	&	2.42	&	12.53 	&	11.30 	&	11.24 	&	15.32 	&	46.46 	\\
Phen$\_$D4+	&	3.23	&	0.18	&	4.35	&	0.72	&	4.11 	&	11.21 	&	16.84 	&	15.05 	&	0.12 	\\
Phen$\_$D9+	&	3.22	&	0.19	&	4.38	&	1.47	&	7.86 	&	10.80 	&	12.15 	&	15.08 	&	15.93 	\\
Pyre$\_$D1+	&	3.23	&	0.24	&	4.38	&	1.59	&	6.59 	&	11.38 	&	7.12 	&	15.87 	&	24.25 	\\
Pyre$\_$D2+	&	3.23	&	0.26	&	4.35	&	0.21	&	0.81 	&	11.47 	&	0.32 	&	15.69 	&	31.48 	\\
Pyre$\_$D4+	&	3.23	&	0.23	&	4.38	&	1.93	&	8.39 	&	11.10 	&	3.11 	&	15.80 	&	7.32 	\\
Pery$\_$D1+	&	3.22	&	0.72	&	4.34	&	1.40	&	1.95 	&	11.06 	&	4.54 	&	17.45 	&	2.12 	\\
Pery$\_$D2+	&	3.22	&	0.68	&	4.36	&	1.96	&	2.88 	&	11.35 	&	0.57 	&	15.19 	&	12.75 	\\
Pery$\_$D3+	&	3.22	&	0.77	&	4.38	&	1.74	&	2.26 	&	11.45 	&	5.11 	&	15.20 	&	2.92 	\\
Coro$\_$D1+	&	3.23	&	1.76	&	4.38	&	2.40	&	1.36 	&	11.33 	&	0.16 	&	14.81 	&	0.15 	\\\hline \noalign{\smallskip}
Average	&	3.22 	&	0.63 	&	4.37 	&	1.38 	&	4.94 	&	11.37 	&	5.56 	&	15.53 	&	18.86 	\\
Stdev	&	0.01 	&	1.01 	&	0.01 	&	0.77 	&	4.47 	&	0.41 	&	7.30 	&	0.67 	&	16.97 	\\
\hline
\noalign{\smallskip} \noalign{\smallskip}
\end{tabular}
\end{center}
\end{table*}
\begin{table*}
\footnotesize
\begin{center}
\caption[]{\footnotesize
           {\it Mean} Intensities of
           the 3.3$\mum$ Aromatic C--H Stretch ($\ACH$) and
           the 4.4$\mum$ Aromatic C--D Stretch ($\ACD$)
           Computed at the {\rm B3LYP/6-311+G$^{\ast\ast}$} Level
           for All the Mono-Deuterated PAHs
           Shown in Figures~\ref{fig:1D_PAH_structures}.
           The $\ACH$ and $\ACD$
           Band Strengths Are on a per C--H or C--D Bond Basis.
           }
\label{tab:MeanBandStrengths}
\begin{tabular}{lcc}
\noalign{\smallskip}\noalign{\smallskip} \hline\hline
Band Strengths	    &	Neutral  PAHs  & Cationic PAHs \\\hline
      $A_{3.3}$ (km$\mol^{-1}$)     &  13.2$\pm{1.0}$      &    0.63$^{+1.01}_{-0.63}$   \\
      $A_{4.4}$ (km$\mol^{-1}$)     &  7.3$\pm{2.4}$       &    1.38$\pm{0.77}$          \\
      $\Aratio$                     &  0.56$\pm{0.19}$     &    4.94$\pm{4.47}$          \\\hline
\noalign{\smallskip} \noalign{\smallskip}
\end{tabular}
\end{center}
\end{table*}

\begin{table*}
\footnotesize
\begin{center}
\caption[]{\footnotesize
                A Summary of the Power Emitted
                from the 3.3$\mum$ Aromatic C--H Band
               $\left(I_{3.3}\right)_{\rm obs}$
               and  the 4.4$\mum$ Aromatic C--D Band
               $\left(I_{4.4}\right)_{\rm obs}$
               Compiled from All the Observational Data
               Available in the Literature.
                }
\label{tab:I44I33_obs}
\begin{tabular}{lcccr}
\noalign{\smallskip} \hline \hline \noalign{\smallskip}
Object	    &  Type   &  	\multicolumn{1}{c}{$\left(I_{3.3}\right)_{\rm obs}^\dag$}
     & \multicolumn{1}{c}{$\left(I_{4.4}\right)_{\rm obs}^\dag$}          &   \multicolumn{1}{c}{$\Iratioobs$}
\\ \noalign{\smallskip} \hline \noalign{\smallskip}
G75.78+0.34$^a$	            &  H{\sc ii} Region  &	22.20	$\pm{	2.00	}$  &	0.30	$\pm{	0.19	}$  &	$1.35\%\pm0.73\%$ \\
NGC\,3603$^a$	            &  H{\sc ii} Region  &	25.10	$\pm{	2.40	}$  &	0.62	$\pm{	0.52	}$  &	$2.47\%\pm1.84\%$ \\
W51 obs.\,1$^a$	            &  H{\sc ii} Region  &	20.00	$\pm{	1.80	}$  &	0.41	$\pm{	0.14	}$  &	$2.05\%\pm0.52\%$ \\
W51 obs.\,2$^a$           	&  H{\sc ii} Region  &	20.60	$\pm{	1.80	}$  &	0.16	$\pm{	0.14	}$  &	$0.78\%\pm0.61\%$ \\
M8$^a$	                    &  H{\sc ii} Region  &	99.10	$\pm{
  8.90	}$  &	1.03	$\pm{	0.53	}$  & $1.04\%\pm0.44\%$ \\
IRAS\,112073 obs.\,1$^a$ 	&  H{\sc ii} Region  &	11.70	$\pm{
  1.00	}$  &	0.30	$\pm{	0.08	}$  & $2.56\%\pm0.46\%$ \\
IRAS\,112073 obs.\,2$^a$ 	&  H{\sc ii} Region  &	12.20	$\pm{
  1.00	}$  &	0.37	$\pm{	0.08	}$  & $3.03\%\pm0.41\%$ \\
Orion Bar$^b$	            &  PDR
&	53.10	$\pm{	0.60	}$  &	1.36	$\pm{	0.11	}$ & $2.56\%\pm0.18\%$ \\
M17$^b$	                    &  PDR
&	14.90	$\pm{	0.20	}$  &	1.40	$\pm{	0.04	}$  & $0.94\%\pm0.26\%$ \\
G18.14.0$^b$	            &  Reflection Nebula
&   4.78	$\pm{	0.06	}$  &	0.12	$\pm{	0.01	}$  & $2.45\%\pm0.20\%$ \\
\hline \noalign{\smallskip}
\end{tabular}
\\
$^\dag$ Flux densities in units of  $10^{-17}$\,W\,m$^{-2}$\,arcsec$^{-2}$     \\
$^a$  Doney et al.\ 2016          \\
$^b$  Onaka et al.\ 2014          \\

\end{center}
\end{table*}


\begin{thebibliography}{30}
\expandafter\ifx\csname natexlab\endcsname\relax\def\natexlab#1{#1}\fi
%

\bibitem[]{}Allamandola, L.J., Tielens, A.G.G.M., \& Barker, J.R.\
            1985, ApJ, 290, L25

\bibitem{}Allamandola, L.~J., Sandford, S.~A., \& Wopenka, B.\
                 1987, Science, 237, 56

\bibitem{}Allamandola, L.J., Tielens, A.G.G.M., \& Barker, J.R.\
            1989, ApJS, 71, 733

\bibitem{}Allamandola, L.J., Hudgins, D.M., \& Sandford, S.A.\
            1999, ApJ, 511, 115

\bibitem{}Bauschlicher, C. W.,  Langhoff, S. R.,
          Sandford, S. A., \& Hudgins, D. M.\ 1997,
          J. Phys. Chem. A, 101, 2414

\bibitem[]{}Bernstein, L.~S., Shroll, R.~M.,
                  Lynch, D.~K., \& Clark, F.~O.\
                  2017, ApJ, 836, 229

\bibitem[]{}Bernstein, M.P., Sandford, S.A., \& Allamandola, L.J.\
            1996, ApJ, 472, L127

\bibitem{}Boesgaard, A. M., \& Steigman, G.\ 1985,
          ARA\&A, 23, 319

\bibitem{}Borowski, P.\ 2012,
            J. Phys. Chem. A, 116, 3866

\bibitem{}Buragohain, M., Pathak, A., Sarre, P., Onaka, T., \& Sakon, I.\ 2015,
          MNRAS, 454, 193

\bibitem[]{}Buragohain, M., Pathak, A., Sarre, P.,
                  Onaka, T., \& Sakon, I.\
                  2016, Planet. Space Sci., 133, 97

\bibitem[]{}Buragohain, M., Pathak, A., Sakon, I.,
                  \& Onaka, T.\ 2020, ApJ, 892, 11

\bibitem{}Burles, S., Nollett, K. M., \& Turner, M. S.\ 2001,
          ApJ, 552, L1

\bibitem{}Coc, A., Vangioni-Flam, E., Descouvemont, P.,
          Adahchour, A., \& Angulo, C.\ 2004,
          ApJ, 600, 544

\bibitem{}Cooke, R., Pettini, M., \& Steidel, Charles C.\ 2018,
          ApJ, 855, 102

\bibitem{}Doney, K. D., Candian, A., Mori, T., Onaka, T.,
          \& Tielens, A. G. G. M.\ 2016,
          A\&A, 586, 65

\bibitem{}Draine, B.~T.\ 2004, 
                in Origin and Evolution of the Elements,
                ed. A. McWilliam \& M. Rauch
                (Cambridge: Cambridge Univ. Press), 317

\bibitem{}Draine, B.~T.\ 2006,
                in Astrophysics in the Far Ultraviolet:
                Five Years of Discovery with FUSE
                (ASP Conf. Ser. 348),  
               ed. G. Sonneborn, H. Moos, \& B.-G. Andersson 
               (San Francisco, CA: ASP), 58

\bibitem{}Draine, B.T., \& Li, A.\ 2001,
          ApJ, 551, 807

\bibitem{}Draine, B.T., \& Li, A.\ 2007,
          ApJ, 657, 810

\bibitem{}Epstein, R. I., Lattimer, J. M., \& Schramm, D. N.\ 1976,
           Nature, 263, 198

\bibitem{}Frisch, M. J., Trucks, G. W., Schlegel, H. B.,
            et al.\ 2009, Gaussian 09, Revision B01,
            Gaussian, Inc., Wallingford CT

\bibitem{}Geballe, T.R., Lacy, J.H., Persson, S.E.,
                McGregor, P. J., \& Soifer, B.T.\
                1985, ApJ, 292, 500

\bibitem{}H{\'e}brard, G., Tripp, T. M., Chayer, P., et al.\ 2005,
          ApJ, 635, 1136

\bibitem{}Hudgins, D. M., Sandford, S. A.,
                Allamandola, L. J.\ 1994,
                J. Phys. Chem., 98, 4243


\bibitem{}Hudgins, D.~M., Bauschlicher, C.~W., Jr.,
                \& Sandford, S.~A.\ 2004,
                ApJ, 614, 770

\bibitem{}Jura, M.\ 1982, in Advances in UV Astronomy,
               Four Years of IUE Research (NASA CP 2238),
               ed. by Y.~Kondo, J.M.~Mead, \& R.D.~Chapman
               (NASA, Greenbelt), 54

\bibitem{}Kwok, S., \& Zhang, Y.\
            2011, Nature, 479, 80

\bibitem{}Li A.\ 2009,
                  in Small Bodies in Planetary Sciences,
                  ed. I.~Mann, A.~Nakamura \& T.~Mukai
                  (Berlin: Springer), 167

\bibitem{}Li, A.\ 2020, Nature Astronomy, 4, 339

\bibitem{}Li, A., \& Draine, B.T.\ 2012,
          ApJ, 760, L35

\bibitem{}Linsky, J. L., Draine, B. T., Moos, H. W., et al.\ 2006,
          ApJ, 647, 1106


\bibitem{}Mazzitelli, I., \& Moretti, M.\ 1980,
          ApJ, 235, 995

\bibitem{}Moos, H. W., Sembach, K. R., Vidal-Madjar, A., et al.\ 2002,
          ApJS, 140, 3

\bibitem{}Onaka, T., Mori, T.~I., Sakon, I.,
                et al.\ 2014,
                ApJ, 780, 114


\bibitem{}Peeters, E., Allamandola, L.~J.,
                 Bauschlicher, C.~W., Jr., et al.\ 2004,
                 ApJ, 604, 252

\bibitem{}Prodanovi{\'c}, T., Steigman, G., \& Fields, B. D.\ 2010,
          MNRAS, 406, 1108


\bibitem{}S{\'a}nchez, Ariel G., Baugh, C. M., Percival, W. J., et al.\ 2006,
          MNRAS, 366, 189

\bibitem[]{}Sandford, S.A.\ 1991, ApJ, 376, 599

\bibitem{}Sandford, S.A., Bernstein, M.P.,
                  \& Dworkin, J.P.\ 2001,
                  Meteoritics Planet. Sci., 36, 1117

\bibitem{}Sandford, S. A., Allamandola, L. J.,
               Tielens, A. G. G. M., et al.\ 1991,
          ApJ, 371, 607

\bibitem{}Sandford, S.A., Bernstein, M.~P.,
                 \& Materese, C.K.\
                2013, ApJS, 205, 8

\bibitem{}Spergel, D. N., Verde, L., Peiris, H. V., et al.\ 2003,
          ApJS, 148, 175

\bibitem{}Steigman, G.\ 2003,
          ApJ, 586, 1120



\bibitem{}Verstraete, L., Puget, J. L., Falgarone, E., et al.\ 1996,
         A\&A, 315, L337

\bibitem{}Wiersma, S.~D., Candian, A., Bakker, J.~M.,
                et al.\ 2020, A\&A, 635, A9

\bibitem{}Wood, B.E., Linsky, J.L., H{\'e}brard, G., et al.\ 2004,
          ApJ, 609, 838

\bibitem{}Yang, X.~J., Glaser, R., Li, A.,
            \& Zhong, J.~X.\
            2013, ApJ, 776, 110

\bibitem{}Yang, X.~J., Glaser, R., Li, A.,
                  \& Zhong, J.~X.\
            2017, New Astron. Rev., 77, 1

\bibitem{}Yang, X.~J., Li, A., \& Glaser, R.\
            2020, ApJS, 247, 1

\bibitem{}Zavarygin, E. O., Webb, J. K.,
              Riemer-S{\o}rensen, S., \& Dumont, V.\ 2018,
              MNRAS, 477, 5536
%
\end{thebibliography}
\end{document}